\documentclass[prd,nofootinbib,twocolumn]{revtex4-1}

\usepackage{mathtools}
\usepackage{amsfonts}
\usepackage{mathrsfs}
\usepackage{enumerate}
\usepackage{color}
\usepackage{graphicx}
\usepackage{paralist}
\usepackage[colorlinks=true]{hyperref}

\begin{document}

\newcommand{\non}{\nonumber}

\title{Minimum energy and the end of the inspiral in the post-Newtonian approximation}
\author{Miriam Cabero}
\email{miriam.cabero@aei.mpg.de}
\author{Alex B. Nielsen}
\author{Andrew P. Lundgren}
\author{Collin D. Capano}
\affiliation{Max Planck Institute for Gravitational Physics (Albert Einstein Institute), Callinstrasse 38, D-30167 Hannover, Germany}
\affiliation{Leibniz Universit\"at Hannover, Welfengarten 1-A, D-30167 Hannover, Germany}

\date{\today}

\begin{abstract}

The early inspiral phase of a compact binary coalescence is well modelled by the post-Newtonian~(PN) approximation to the orbital energy and gravitational wave flux. The transition from the inspiral phase to the plunge can be defined by the minimum energy circular orbit~(MECO). In the extreme mass-ratio limit the PN energy equals the energy of the (post-Newtonian expanded) exact Kerr solution. However, for comparable-mass systems the MECO of the PN energy does not exist when bodies have large spins and no analytical solution to the end of the inspiral is known. By including the exact Kerr limit, we extract a well-defined minimum of the orbital energy beyond which the plunge or merger occurs. We study the hybrid condition for a number of cases of both black hole and neutron stars and compare to other commonly employed definitions. Our method can be used for any known order of the post-Newtonian series and enables the MECO condition to be used to define the end of the inspiral phase for highly spinning, comparable mass systems.
\end{abstract}

\maketitle

\section{Introduction}

Binary systems are ubiquitous in the universe. While most binary systems are well described by Newtonian physics, some systems orbit closely enough that relativistic effects become important for describing their dynamics. Systems involving compact objects such as black holes and neutron stars are able to orbit very close together and have been observed both with radio observations~\cite{Lorimer:2008se,Kramer:2009zza} and, recently, by the Advanced Laser Interferometer Gravitational-wave Observatories (Advanced LIGO)~\cite{GW150914, GW151226, O1BBH}. The full solution to the two body problem describing these dynamics has not been solved analytically in general relativity. If the mass of the secondary body can be neglected then the orbit can be approximated as a test-mass orbit in a one-body solution, such as the Kerr solution, which describes a rotating black hole~\cite{Kerr:1963ud}. In this case there is an innermost equatorial orbit beyond which the test-mass cannot orbit equatorially. This innermost stable circular orbit (ISCO) is often taken as the inner-edge of the accretion disks around black holes~\cite{BHastrophysics, SpinAccretion}. A description of the system by test-mass dynamics, however, is meaningful only when the mass of the companion is much smaller than the mass of the primary object.

If the companion is another object of comparable mass, the spacetime of the system is not well described by the Kerr solution. A number of approximation methods are used in this case, including the post-Newtonian approximation. Post-Newtonian (PN) theory yields an analytical approximation of the motion of compact binaries, under the assumptions of weak gravitational fields inside the sources and of slow internal motion~\cite{PN2body}. Despite these limitations, the PN approximation has proven to be an unexpectedly effective description~\cite{CliffWill}. Observational limits on deviations from the calculated PN coefficients are given in~\cite{O1BBH}.

The radiation of energy in the form of gravitational waves plays an important role in the motion of relativistic compact binaries. The gravitational-wave emission causes the orbit to gradually shrink, bringing the bodies closer together in a long inspiral phase~\cite{PetersMathews}. Evidence for this effect has been observed in pulsar systems such as the Hulse-Taylor binary pulsar~\cite{HulseTaylor2004} or the ``double-pulsar" PSR J0737-3039~\cite{2hour}, and also in the binary black holes detected with Advanced LIGO~\cite{GW150914, GW151226, O1BBH}. After the long inspiral, the evolution is typically followed by a plunge, merger, and ringdown. However, these processes can be interrupted if one of the objects (or both) is a neutron star. Depending on its internal composition and orbital parameters, the neutron star can be tidally deformed and even completely disrupted before the plunge phase~\cite{GW_eos, tidal_disruption}. This has been proposed as a possible mechanism to explain short gamma-ray bursts~\cite{Eichler:1989ve}.

In the absence of tidal disruption, the orbital energy gradually decreases until it reaches a minimum. Based on physical insight from the extreme mass-ratio limit in the Kerr spacetime, this minimum, when it exists, can be considered as the end of the inspiral. The orbit at which the energy reaches its minimum value is called the minimum energy circular orbit (MECO)\footnote{Also innermost circular orbit (ICO) in the literature~\cite{BlanchetICO}.}. If the PN approximation for the energy is valid, the MECO will be the minimum of the PN energy. As we will see, direct application of the minimum energy condition to the known PN energy leads to a MECO that depends sensitively on the PN order and the intrinsic angular momentum (spin) of the black hole.

In the Schwarzschild and Kerr spacetimes the MECO coincides with the ISCO, as well as in any system defined by an exact Hamiltonian~\cite{BuonannoChenVallisneri, BarausseSelfForce}. However, it has been shown that this statement is not necessarily true in the PN approximation~\cite{CB-dynamics, BuonannoChenVallisneri, BlanchetICO}. Blanchet and Iyer~\cite{CB-dynamics} computed the ISCO for non-spinning objects in the PN formalism studying the stability of circular orbits against linear perturbations of the equations of motion. They observed that the PN corrections increase the frequency of the ISCO with respect to the Schwarzschild solution and thus the ISCO radius is smaller in PN theory than in the Schwarzschild case. A generalisation of their method for spinning objects has been performed by Favata~\cite{iscoFavata}. 

Models for the complete evolution of binary black hole systems have recently been developed~\cite{SEOBNRv4, Khan:2015jqa}. These extend PN techniques by including numerical relativity results from regions where the PN approximation breaks down. However, the end of the inspiral in compact binaries is still of great theoretical and practical interest for the two-body problem in general relativity. The dynamics of the coalescence qualitatively change in the transition from the inspiral to the plunge. PN theory breaks down close to the merger due the nature of the approximation. This breakdown is not well-defined, yet a critical issue for gravitational-wave modelling.  

The merger represents the violent collision of two compact objects in the fully non-linear regime. The parameters of these objects must be independently measured prior to merger in order to test whether the object that results from such a collision is compatible with what general relativity predicts. For instance, some tests of general relativity with gravitational-wave observations distinguish between the inspiral and the last phases of the waveform~\cite{TestsGR150914, GoldenIMR}, and tests of Hawking's area theorem~\cite{AreaTheorem} can be restricted to the inspiral phase when determining the initial areas~\cite{AreaTheoremTest}. Therefore, waveforms to describe the inspiral phase require knowledge of a suitable end point. In the past, either the exact Schwarzschild ISCO or the PN MECO have been used as the frequency cutoff for inspiral waveforms (see~\cite{TitoSpin, BIOPS, T2T4}, for instance). However, both approaches have their limitations. The existence of an ISCO in the full two-body problem is uncertain and hard to calculate in the PN framework. The validity of using the test-mass value is also not guaranteed for comparable-mass systems. Conversely, the MECO does not always have a finite value. This limits the PN terms that can be included when using this method.

In this work we study the properties of the MECO in the post-Newtonian theory. We seek to obtain a MECO that exists for any known PN order, any mass-ratio and any value of the spins of the objects. In the extreme mass-ratio inspiral case, the PN approximation is poorly convergent, but one can use the test-mass dynamics of the exact solution directly~\cite{BlanchetICO}. Following the idea of a hybrid approach introduced in~\cite{Inspiral-to-plunge} we include exact test-limit results into the PN approximation and show that it fulfils our criteria. In~\cite{Inspiral-to-plunge} this was used only with non-spinning Schwarzschild like systems, but we extend this to spinning systems. In our case, the MECO represents the maximum limit of the validity of the PN approximation. We compare our limit to the point of peak amplitude emission in the spinning effective-one-body model calibrated to numerical relativity (SEOBNR~\cite{SEOBNRv4}). The SEOBNR peak is the expected value of the instantaneous gravitational-wave frequency at the time when the (2,2) mode amplitude peaks in numerical simulations. Typically, the time at which the common apparent horizon forms in numerical simulations of binary black hole mergers is very close to the time at which the amplitude peaks. Therefore, the SEOBNR peak can be viewed as a proxy for the frequency at which the merger occurs~\cite{Taracchini_private}. If there is a plunge phase, then the end of the inspiral will occur before this.

We are particularly interested here in binaries with equal masses and low mass-ratios, where PN is expected to be better behaved, but no exact solution is known. For comparison, we consider mainly two cases: 
\begin{inparaenum}[i)]
\item a neutron-star black-hole binary (NSBH) of mass-ratio $q = m_\textrm{BH}/m_\textrm{NS} \simeq 7$, where the spin of the neutron star is set to zero, and
\item an equal-mass binary black hole (BBH), where both black holes have the same spin.
\end{inparaenum}
For these binary systems, we analyse in section~\ref{MECO} the dependence of the MECO on the spin and on the PN order. In section~\ref{TestMass}, we use the exact Kerr MECO to analyse the test-mass limit. Section~\ref{Corrected_MECO} computes a well-defined MECO which has a finite value for any PN order and any spin. Since our result depends only on the mass-ratio and not on the total mass of the binary, it can also be used for systems containing intermediate-mass or supermassive black holes. In section~\ref{SNRgain} we use the signal-to-noise ratio (SNR) as a measure of approximately how much extra inspiral signal is gained or lost when using different choices for the end of the inspiral. Section~\ref{Approximants} investigates if the integrands used for PN waveform generation are well-posed through the new MECO termination point. This result is important for faithfulness studies of different waveform models. Finally, as an example of the flexibility of our approach, section~\ref{Tidal} considers tidal effects, such as those published in~\cite{TidalEffects,TidalEffects2}, in NSBH with comparable masses (mass-ratio $q \simeq 2$) and shows how they can be included straightforwardly. 

Throughout this paper we use geometrical units $G = c = 1$. We present most of our results in terms of the PN velocity parameter $v$. In the Newtonian limit this corresponds to the sum of the orbital speeds of the two orbiting objects and also in the Schwarzschild limit corresponds to the Schwarzschild coordinate velocity of an orbiting test mass. However, in more general cases, $v$ should be interpreted solely as a formal expansion parameter related to the gravitational wave dominant mode frequency, $f_{\text{GW}}$, by $v^3 = \pi M f_{\text{GW}}$. We denote the total mass by $M = m_1 + m_2$, the individual masses of the bodies by $m_i$, the mass-ratio by $q = m_1/m_2 \geq 1$ and the symmetric mass-ratio by $\eta = m_1 m_2 / (m_1 + m_2)^2 \leq 0.25$. For simplicity, we consider the case where the black-hole spin is aligned or anti-aligned with the orbital angular momentum. Precessing systems are expected to broadly follow similar lines in terms of their projected spins, but we leave a detailed analysis to further work. The projection of the dimensionless spin of the black hole onto the orbital angular momentum is denoted by $\chi_i = \vec{S}_i \cdot \hat{L} / m_i^2$, where $\vec{S}_i$ is the intrinsic angular momentum and $\hat{L}$ is the unit vector along the orbital angular momentum. The projected spin can take values $-1 \leq \chi_i \leq 1$, where positive spins indicate alignment with the orbital angular momentum, and negative spins indicate anti-alignment.

% -------------------------------------------------------------------------------------------------------------------------------

\section{Behaviour of the MECO at different post-Newtonian orders}\label{MECO}

The PN energy and flux are given as series expansions in the PN velocity parameter, $v$. Terms of order $v^{2n}$ are called $n$PN terms, where even (odd) powers of $v$ have integer (non-integer) $n$. Under the assumption of circular orbits, the PN expressions get considerably simplified (see Appendix~\ref{PNformulae}). This is a reasonable approximation for late-time systems dominated by gravitational wave emission, since the decay of the orbital eccentricity happens much faster than the coalescence of isolated circularly orbiting binaries~\cite{Peters}. For non-spinning systems, the PN energy in the centre-of-mass frame for circular orbits is known up to $4$PN order and the flux up to $3.5$PN~\cite{BlanchetPN2014,4PN}. The spin corrections to the energy are known up to the 4PN spin-spin~\cite{2PNspin-spin, 3PNspin-spin, LeviSteinhoff_3PNspin-spin, LeviSteinhoff_4PNspin-spin} and the $3.5$PN spin-orbit~\cite{spin-orbit, LeviSteinhoff_spin-orbit} terms. Spin-cubed terms appear at the $3.5$PN order~\cite{spin-cubed, LeviSteinhoff_spin-orbit}, and spin-quartic terms at $4$PN order~\cite{LeviSteinhoff_4PNspin-spin}. 

For consistency, the PN energy in the test-mass limit ($m_2 \to 0$, $m_1$ fixed) should reproduce the Taylor expansion of the orbital energy of a test-mass in the Kerr spacetime up to the PN order considered. In this extreme mass-ratio limit, one can use the exact Kerr solution, where the end of the inspiral is given by the Kerr ISCO. The location of the Kerr ISCO in the equatorial plane depends on the spin of the black hole and, in Boyer-Lindquist coordinates, is given by~\cite{KerrISCO}
\begin{align}
r = {} & m \left[ 3 + Z_2 \mp \sqrt{ (3 - Z_1) (3 + Z_1 + 2Z_2)} \right] \, , \label{Kisco} \\
\text{where} \quad & Z_1 = 1 + \left( 1 - \chi^2 \right)^{1/3} \left[ \left( 1 + \chi \right)^{1/3} + \left( 1 - \chi \right)^{1/3} \right] , \non \\ 
& Z_2 = \sqrt{ 3 \chi^2 + Z_1^2} \, , \non
\end{align}
$m$ is the mass of the Kerr black hole, and $\chi$ is its spin relative to the orbital angular momentum. For spin zero, the Kerr ISCO reduces to the Schwarzschild ISCO, $r = 6 m$. The upper sign in Eq.~\eqref{Kisco} is for prograde orbits (spin of the black hole aligned with the orbital angular momentum), while the lower sign is for retrograde orbits (spin anti-aligned with the orbital angular momentum). For extremal spin ($\chi = \pm 1$), the Kerr ISCO is located at $r = m$ (aligned case) and $r = 9m$ (anti-aligned case). If the mass and spin of the smaller body are not totally neglected, but considered small, self-force calculations provide the corrections due to the effect of the second body~\cite{Self-force, SelfForcetoSchwarzschild, SelfForcetoKerr}.

\begin{figure}[tb]
\centering
	\includegraphics[width=\columnwidth]{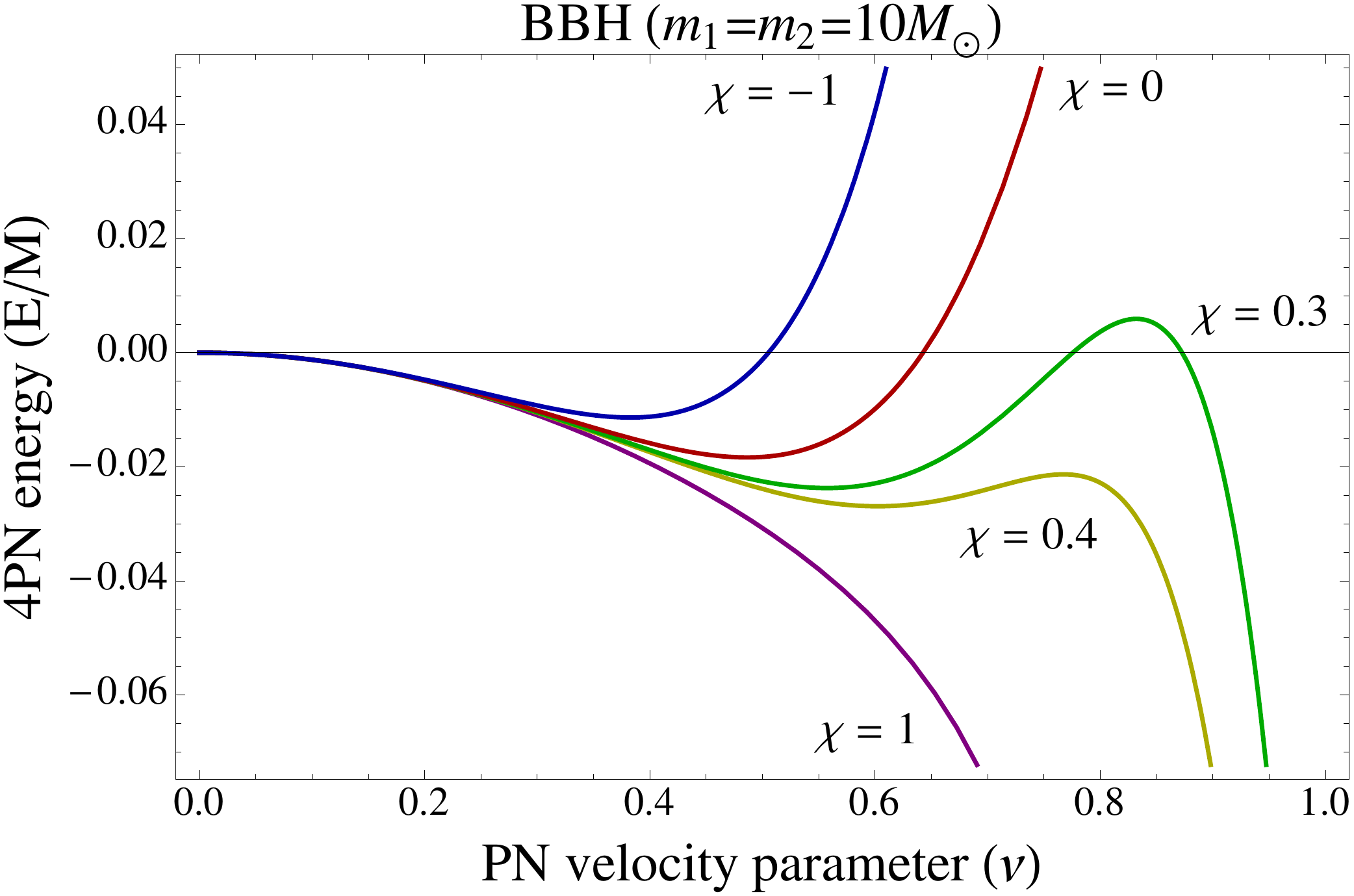}
\caption{Current 4PN energy per unit mass as function of the PN velocity parameter, $v$, for an equal-mass BBH. The two black holes are considered to have equal spin, $\chi_1 = \chi_2 = \chi$. For some values of positive spins the energy shows a maximum after the minimum, as can be seen with $\chi \simeq 0.3$ for instance. For higher spins, the energy does not reach a minimum and therefore the MECO does not exist at the 4PN order.}
\label{Energy}
\end{figure}

As the mass-ratio decreases, the system does not follow test-mass dynamics of the exact Kerr solution. The PN approximation can be used in this regime to calculate a PN orbital energy. The MECO is then defined as the minimum of this PN energy, corresponding to the velocities at which $dE^\textrm{PN}/dv=0$. Figure~\ref{Energy} shows the energy including terms up to 4PN order for BBH systems with different spins, where in each case both black holes have the same spin value. The orbital energy decreases and reaches a minimum that represents the end of the inspiral (the MECO). What occurs after the minimum is to be discarded physically, since the adiabatic assumption is certainly violated beyond this point. For high spin values, the energy does not show a minimum, and therefore the MECO does not exist. Due to the absence of a minimum of the energy in certain regions of the parameter space, the current MECO cannot be used as a robust definition of the end of the inspiral.

\begin{figure}[tb]
\centering
	\includegraphics[width=\columnwidth]{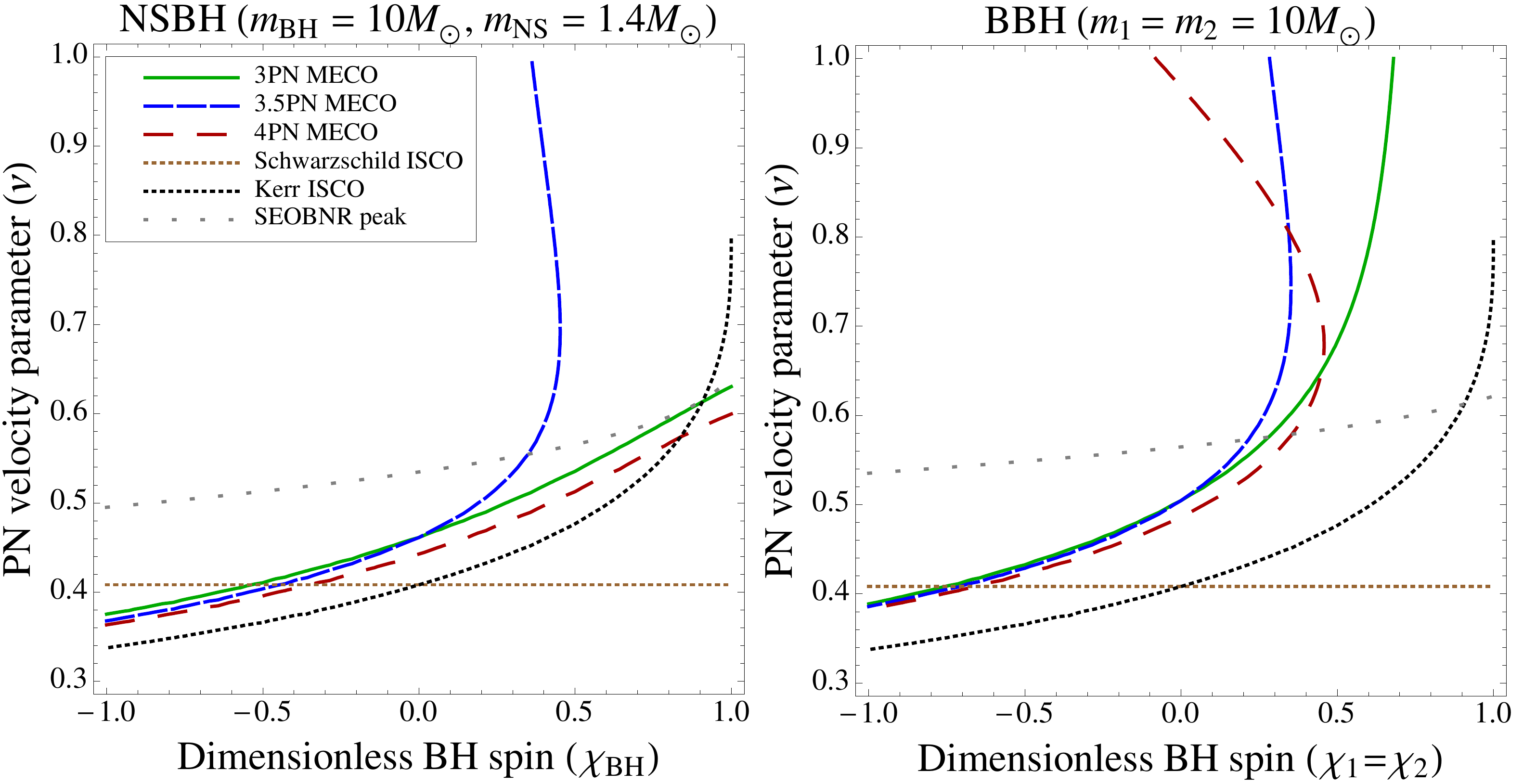}
\caption{Different inspiral-cutoffs for NSBH and BBH as function of the PN velocity parameter, $v$. Dotted lines represent the Schwarzschild ISCO (which is independent of the spin value), the Kerr ISCO, and the SEOBNR peak. Continuous and dashed lines represent the extrema of the PN energy (the MECO and possible maxima of the energy) up to different PN orders.  The neutron-star spin is considered negligible compared to the black-hole spin and therefore set to zero. The black-hole spins are equal in the BBH case and aligned (positive $\chi$) or anti-aligned (negative $\chi$) with the orbital angular momentum. Except for the even 3PN and 4PN orders in the NSBH case, the MECO does not exist in the region of high aligned spins.}
\label{MECOs}
\end{figure}

Figure~\ref{MECOs} shows the PN velocity parameter of the PN MECO when including terms up to three different PN orders: 3PN, 3.5PN and 4PN. The double valued curves for the 3.5PN and the 4PN BBH cases correspond to maxima and minima, as shown for example in the $\chi = 0.3$ curve of Fig.~\ref{Energy}. For comparison, the ISCOs (see Appendix~\ref{Relations} for the relations between the ISCO radius and its orbital velocity) and the SEOBNR peak (obtained from the LSC Algorithm Library Suite~\cite{LALSuite}) are also shown. As can be seen in Fig.~\ref{MECOs}, the MECO depends sensitively on the PN order, and does not exist in certain regions of parameter space for certain orders. For high values of the black-hole spins in the BBH case for instance, the MECO does not exist for any of the three PN orders shown. The NSBH shows the same behaviour for the odd 3.5PN order, while the even 3PN and 4PN energies reach a minimum for any value of the black-hole spin. 

% -------------------------------------------------------------------------------------------------------------------------------

\section{PN expansions of the test-mass limit}\label{TestMass}

The PN approximation is poorly convergent in the test-mass limit, but is expected to be more accurate in the comparable-mass regime~\cite{BlanchetICO}. Therefore, one would expect the MECO to be well-defined for comparable-mass binaries. However, the previous section has shown that the MECO does not exist for high spins in the equal-mass case. Since this issue does not seem to be mitigated by the addition of higher PN orders, here we analyse the test-mass case, where the PN approximation is already known to fail.

The orbital energy per unit mass of a test-mass around a Kerr black hole can be given as~\cite{KerrISCO, iscoFavata}
\begin{equation} \label{KerrEnergy}
E^\textrm{Kerr} = \left( \frac{1 - 2 w + \chi w^{3/2}}{\sqrt{1 - 3 w + 2 \chi w^{3/2}}} - 1 \right) \, ,
\end{equation}
where $w = \frac{v^2}{(1 - \chi v^3)^{2/3}}$. Here, $\chi$ is the projected spin of the black hole, and $v$ is the equivalent PN velocity parameter of the test-mass particle given by $v^3 = 2\pi M / T$, where $T$ is the period of the orbit in Boyer-Lindquist coordinates.
For a Kerr black hole, the ISCO coincides with the MECO, which is obtained by solving
\begin{equation}
\frac{d E^\textrm{Kerr}}{dv} = 0 \, .
\end{equation}
This reduces to obtaining the velocity at which
\begin{align} \label{KerrMECO}
\Big[ 3 \chi^2 v^4 - \left( 1 + 7 \chi v^3 \right) & \left( 1 - \chi v^3 \right)^{1/3} \non  \\
& + 6 v^2 \left( 1 - \chi v^3 \right)^{2/3} \Big] = 0 \, .
\end{align}

In the test-mass limit ($m_2 \to 0$, $m_1$ fixed), the PN energy [given in Appendix \ref{PNformulae}] equals the Taylor expansion of the Kerr energy up to the highest PN order known. Expanding Eq.~\eqref{KerrEnergy} up to 4PN, we obtain~\footnote{The dimensionless spin~$\chi$ in Eq.~\eqref{TaylorKerr} and the PN spin parameter~$S$ used in Appendix~\ref{PNformulae} are related through $\chi = S / m^2$.}
\begin{align} \label{TaylorKerr}
E^\text{Kerr} (\chi_1) \simeq {}& \frac{E^\text{PN}}{\eta} \bigg |_{m_2 \to 0} \non \\
= {}& - \frac{1}{2} v^2 \bigg[ 1 - \frac{3}{4} v^2 + \frac{8 \chi_{1}}{3} v^3 - \left( \frac{27}{8} + \chi_{1}^2 \right) v^4 \non \\
& + 8 \chi_{1} v^5 - \left( \frac{675}{64} + \frac{65 \chi_{1}^2}{18} \right) v^6 + 27 \chi_{1} v^7 \non \\
& - \left( \frac{3969}{128} + \frac{469 \chi_{1}^2}{24} \right) v^8 + \mathcal{O} \left(v^9\right) \bigg] \, .
\end{align}

\begin{figure}[tb]
\centering
	\includegraphics[width=0.49\columnwidth]{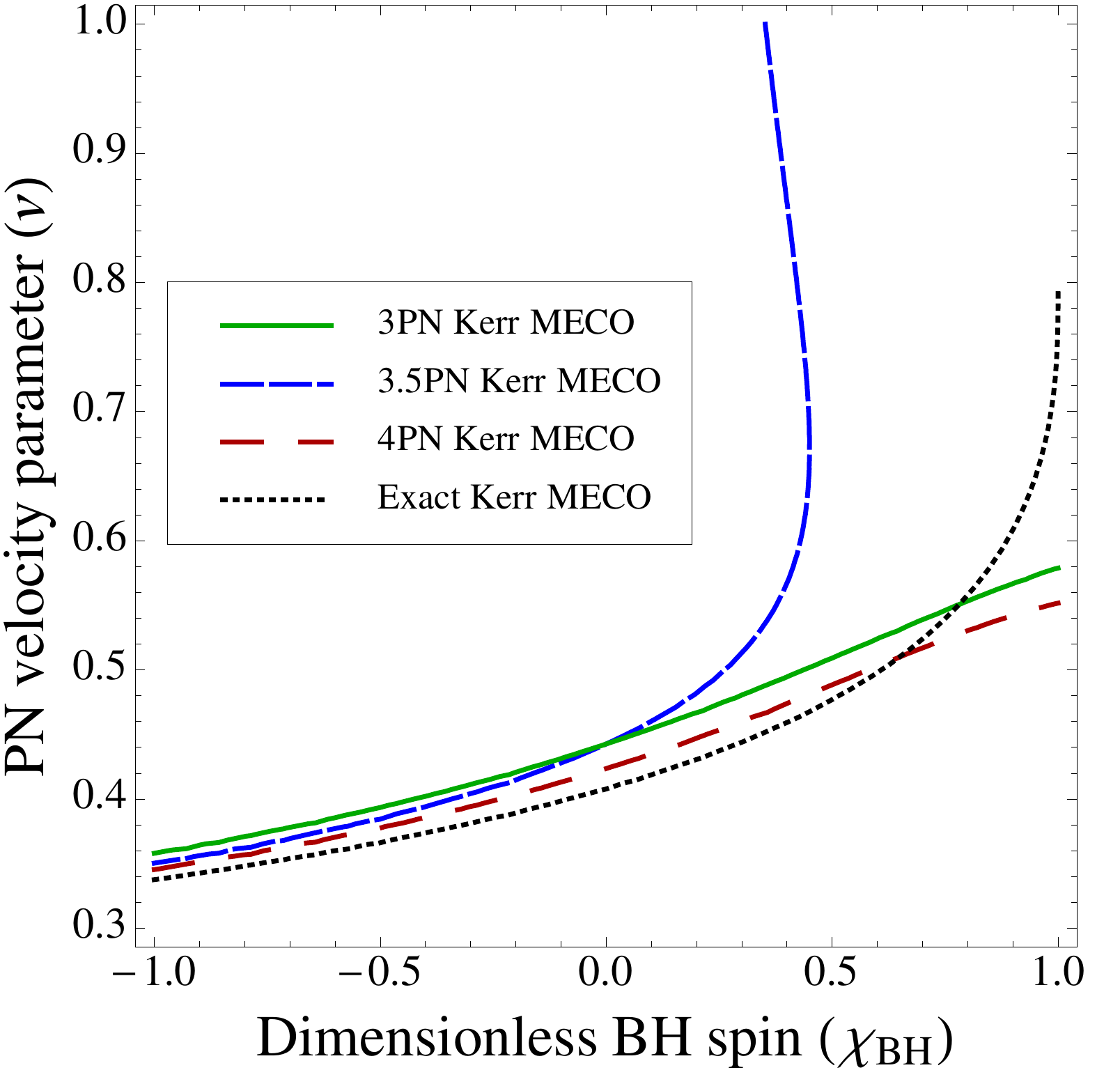}
	\includegraphics[width=0.49\columnwidth]{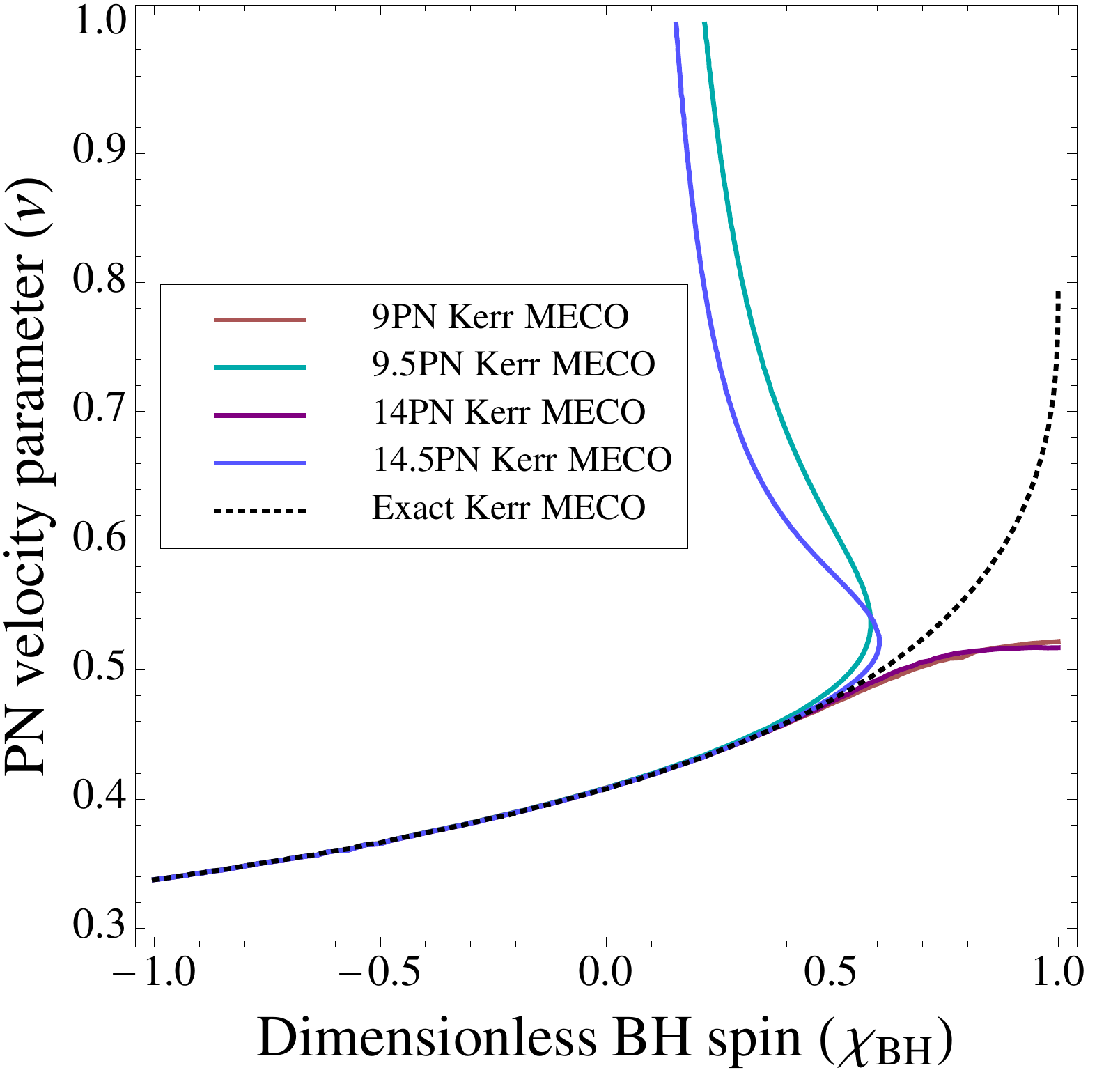}
\caption{MECO for a test mass orbiting a Kerr black hole as function of the PN velocity parameter, $v$. The dotted line represents the exact solution. Continuous and dashed lines show different orders of the approximation~\eqref{TaylorKerr}, which are called ``PN orders'' for comparison with the PN approximation. The expanded MECO does not approach the exact MECO at high spins, showing a very unstable behaviour for odd PN orders (the upper branch of odd PN orders corresponds actually to a maximum of the energy). Therefore, any attempt to obtain a well-defined MECO for the PN energy at odd orders will fail with the current method.}
\label{KerrMECOs}
\end{figure}

A comparison of the exact Kerr MECO [Eq.~\eqref{KerrMECO}] to its approximation at different ``PN orders'' (given by the minimum of Eq.~\eqref{TaylorKerr} up to the chosen order) can be seen in Fig.~\ref{KerrMECOs}. At higher spins, the MECO of the 4PN expansion for the Kerr energy does not resemble the MECO of the exact Kerr energy. Therefore, the MECO of the post-Newtonian energy is unlikely to resemble the MECO of the unknown exact solution, if it exists. Figure~\ref{KerrMECOs} also shows that even if one knew the post-Newtonian energy up to 14.5PN, the test-mass part would likely only agree with the exact known solution up to spins of $\chi \simeq 0.5$, or velocities of $v \simeq 0.45$. Another remarkable feature of the expansion of the Kerr energy is the different behaviour between odd and even PN orders. While even PN orders appear to have a minimum for any value of the spin, odd PN orders do not present extrema for $\chi \gtrsim 0.5$. This behaviour is very similar to what we have seen for the complete PN energy in the NSBH case (only one spinning object). 

% -------------------------------------------------------------------------------------------------------------------------------

\section{Stabilisation of the MECO}\label{Corrected_MECO}

Mathematically, it is unknown if the PN series can be derived as a Taylor expansion of a family of exact solutions~\cite{PN2body}. However, for consistency, its test-mass limit should always equal the expansion of the exact Kerr solution. This expansion has been shown to have a very unstable MECO that is often far away from the exact Kerr ISCO. It is necessary to avoid the test-mass truncation in order to obtain a stable MECO which is well-defined for any PN order and any spin value. In this section we extend an idea in~\cite{Inspiral-to-plunge} of a hybrid energy to the case of spinning objects, to obtain a well-defined MECO condition for all spin values.

Consider a hybrid energy given by replacing the test-mass expansion in the PN energy (up to the PN order known) with the exact Kerr solution:
\begin{align} \label{NewEnergy}
E^{\text{h}} ={}& \frac{E^{n\text{-PN}}}{\eta} - \left( \sum_{x=0}^{x=2n} E^{\text{Kerr}} (v^x) \right) + E^{\text{Kerr}}  \non \\
={}& E^{\text{Kerr}}  - \frac{1}{2} v^2 \left\{ - \frac{\eta}{12} v^2 + \left( \frac{19}{8} \eta - \frac{\eta^2}{24} \right) v^4 \right. \non \\
& + \left[ \left( \frac{34445}{576} - \frac{205 \pi^2}{96} \right) \eta - \frac{155}{96} \eta^2 - \frac{35}{5184} \eta^3 \right] v^6  \non \\
& - \bigg[ \left( \frac{123671}{5760} - \frac{9037 \pi^2}{1536} - \frac{1792}{15}  \ln 2 - \frac{896}{15} \gamma_{E} \right) \eta \non \\
& \quad + \left( \frac{498449}{3456} - \frac{3157 \pi^2}{576} \right) \eta^2 - \frac{301}{1728} \eta^3 \non \\
& \quad - \frac{77}{31104} \eta^4 - \frac{448}{15} \eta \ln v^2 \bigg] v^8 \non \\
& + \left[ E_\textrm{SO} - \chi \left( \frac{8}{3} v^3 + 8 v^5 + 27 v^7 \right) \right] \non \\
&  \left. + \left[ E_\textrm{SS} + \chi^2 v^4 + \frac{65}{18} \chi^2 v^6 + \frac{469}{24} \chi^2 v^8 \right] \right\} \, ,
\end{align}
where $E^\textrm{Kerr}$ is the energy given in Eq.~\eqref{KerrEnergy}, $n$ is the PN order chosen, and $E_\textrm{SO}$, $E_\textrm{SS} = E_\textrm{SS}^\textrm{2PN} + E_\textrm{SS}^\textrm{3PN} + E_\textrm{SS}^\textrm{4PN}$ are the spinning terms of the PN energy given in Appendix~\ref{PNformulae} (recall the relation $S_i = \chi_i m_i^2$). Here we do not include the spin-cubic term because its effect is unnoticeable for our purposes. The spin-quartic term vanishes for the systems we are considering~\eqref{S4}.

When including single-body Kerr terms in a two-body energy, the choice of the spin $\chi$ that comes from the Kerr energy is not unique. Ideally, one would want a spin that keeps the energy symmetric under exchange $1 \leftrightarrow 2$. Motivated by~\cite{EffectiveSpinAjith, EffectiveSpinSantamaria}, we use the mass-weighted effective spin parameter introduced in~\cite{EffectiveSpinDamour}
\begin{equation}
\chi_{\text{eff}} = \frac{\chi_1 m_1 + \chi_2 m_2}{m_1 + m_2} \, .
\end{equation}
This effective spin parameter reduces to a single black hole spin in the test-mass limit, and is typically used in gravitational-wave astronomy~\cite{O1BBH}.

\begin{figure}[tb]
\centering
	\includegraphics[width=\columnwidth]{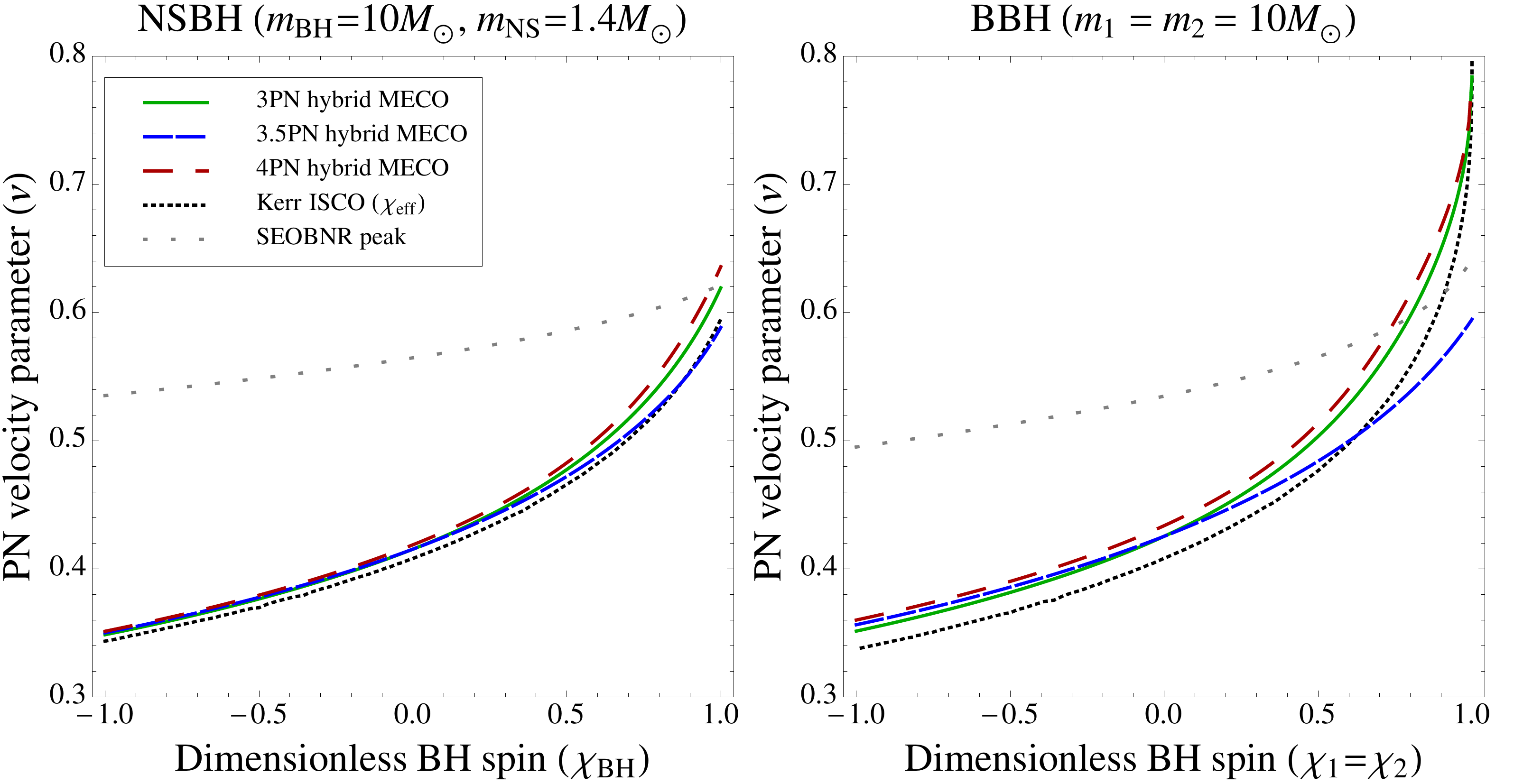}
\caption{MECO of the hybrid energy given in Eq.~\eqref{NewEnergy} as function of the PN velocity parameter, $v$. Dotted lines represent the Kerr ISCO, with $\chi = \chi_{\text{eff}}$, and the SEOBNR peak. Continuous and dashed lines represent the minimum of the hybrid energy up to different PN orders. There is still a clear difference between odd and even PN orders at high aligned spins. However, the energy always reaches a minimum, suggesting that the hybrid MECO is a better description of the end of the inspiral than the MECO of the pure PN energy.}
\label{CorrectedMECO}
\end{figure}

The hybrid energy in Eq.~\eqref{NewEnergy} has a minimum for any known PN order and for any spin value, as can be seen in Fig.~\ref{CorrectedMECO}. There is still a difference between PN orders, although the 3.5PN spin-orbit term seems to push the minimum to lower velocities with respect to the energies up to even PN orders in the region of positive spins. In the absence of higher order spinning terms, one cannot conclude if spin-orbit terms will always have this effect on the hybrid MECO. Nevertheless, due to the stability of this approach at any known PN order and at any spin value, we suggest the use of the hybrid MECO as a PN approximation to the end of the inspiral phase. More specifically, we suggest the use of the 3.5PN hybrid MECO, which is more conservative than the even orders at higher spins, and, for the binary systems considered here, is always below the SEOBNR peak.

\begin{figure}[tb]
\centering
	\includegraphics[width=\columnwidth]{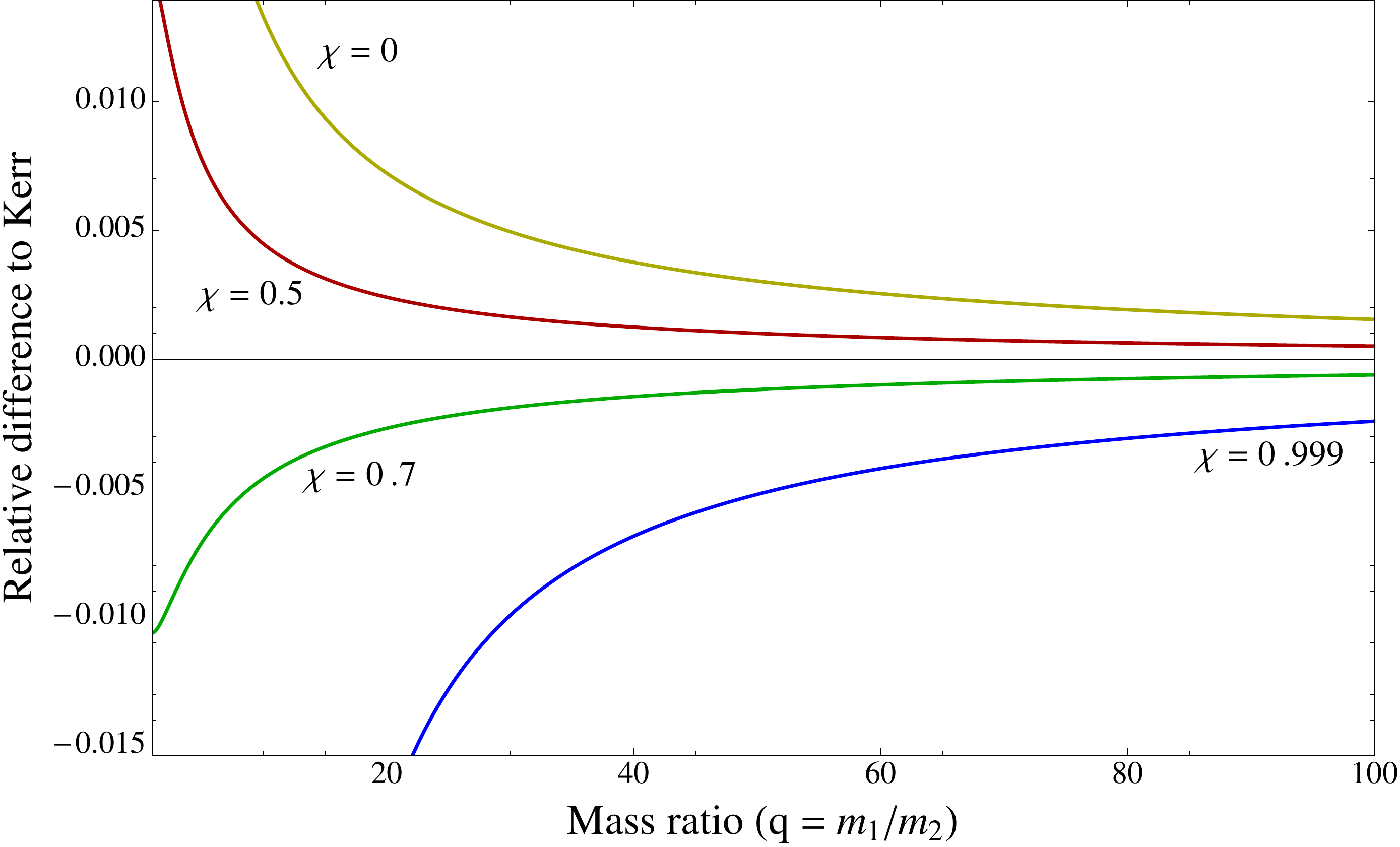} \\
	\includegraphics[width=\columnwidth]{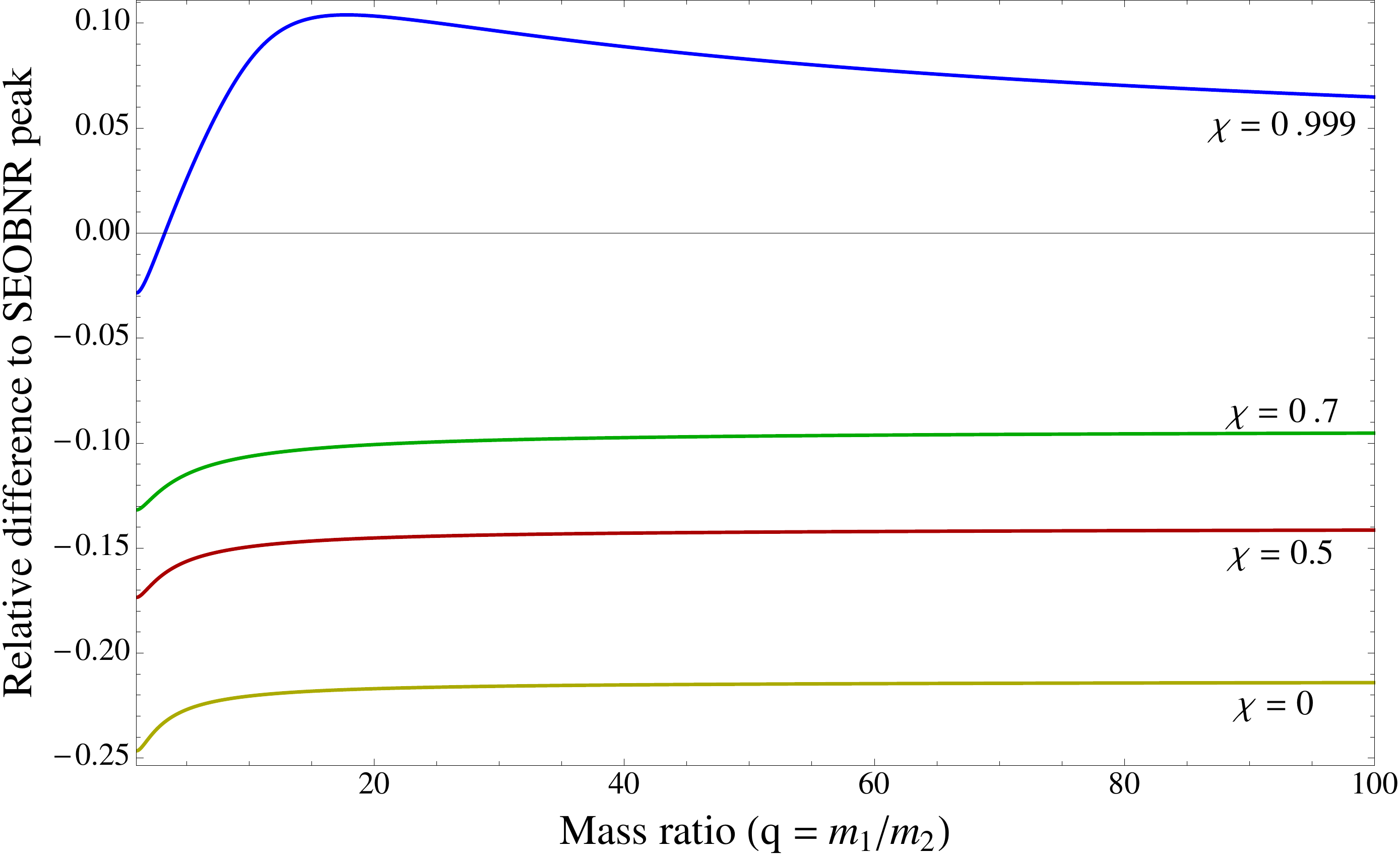}
\caption{Relative difference, for BBH with mass-ratios up to $q = 100$, between the PN velocity parameter of the hybrid MECO at 3.5PN order and (\textit{top}) of the exact Kerr MECO, $(v_\textrm{hybrid} -~v_\textrm{Kerr}) / v_\textrm{Kerr}$, and (\textit{bottom}) of the SEOBNR peak, $(v_\textrm{hybrid} - v_\textrm{SEOB}) / v_\textrm{SEOB}$. The two black holes are considered to have equal spin, $\chi_1 = \chi_2 = \chi$. Only aligned spins (positive) are shown, because anti-aligned spins behave similarly to the spin zero case.}
\label{Hybrid_diff}
\end{figure}

The relative difference in the velocity between the exact Kerr MECO and this new hybrid MECO at 3.5PN order is shown in the top plot of Fig.~\ref{Hybrid_diff} as a function of the mass-ratio. The deviation from the exact Kerr ISCO depends on the mass-ratio: increasing the mass-ratio leads to higher similarity with the exact Kerr ISCO. Therefore, the difference tends to zero asymptotically. The bottom plot of Fig.~\ref{Hybrid_diff} shows the relative difference in the velocity between the SEOBNR peak and the hybrid MECO. For spins $\chi \gtrsim 0.94$, the hybrid MECO lies above the SEOBNR peak when the mass ratio is increased. However, we do not consider this a critical defect of our model, since those parameters lie outside the calibration region of the SEOBNR model.

% -------------------------------------------------------------------------------------------------------------------------------

\section{Relative SNR between inspiral templates} \label{SNRgain}

We have seen how the hybrid MECO termination condition differs from the Schwarzschild ISCO and other conditions in terms of the PN velocity parameter $v$. This velocity parameter is simply related to the orbital frequency and the observed instantaneous frequency of emitted gravitational waves. To evaluate how significant the difference between terminations is in terms of the sensitivity of ground-based gravitational wave detectors, we turn now to the question of how much of the actual inspiral signal is contained between the Schwarzschild ISCO and the hybrid MECO.
This is analogous to computing the relative signal-to-noise ratio (relative SNR) between a template waveform that terminates at the hybrid MECO frequency~($f_\textrm{hM}$), and a template waveform that terminates at the Schwarzschild ISCO frequency ($f_\textrm{Schw}$).

\begin{figure*}[tb]
\centering
	\includegraphics[width=0.4\textwidth]{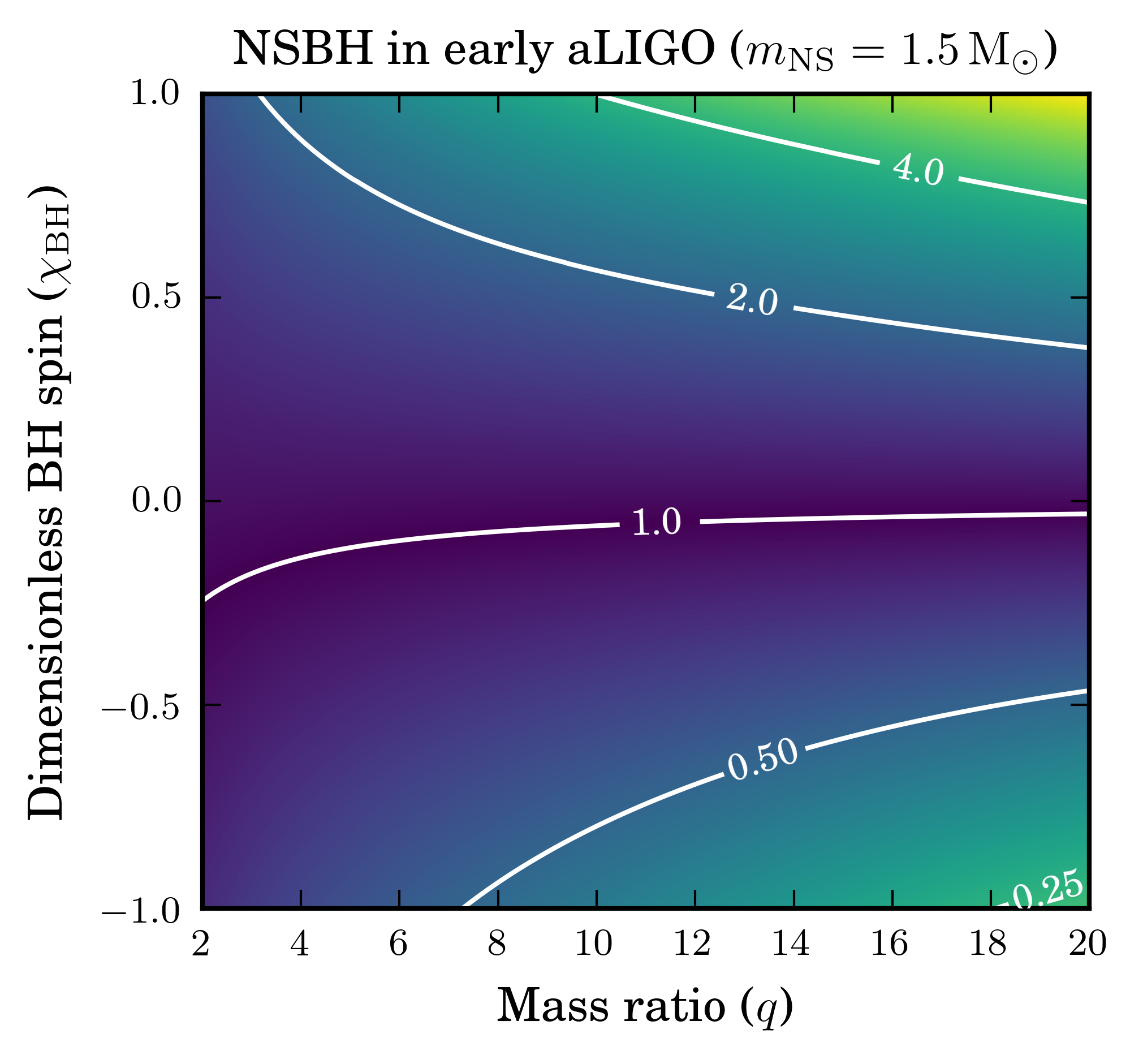} \qquad
	\includegraphics[width=0.4\textwidth]{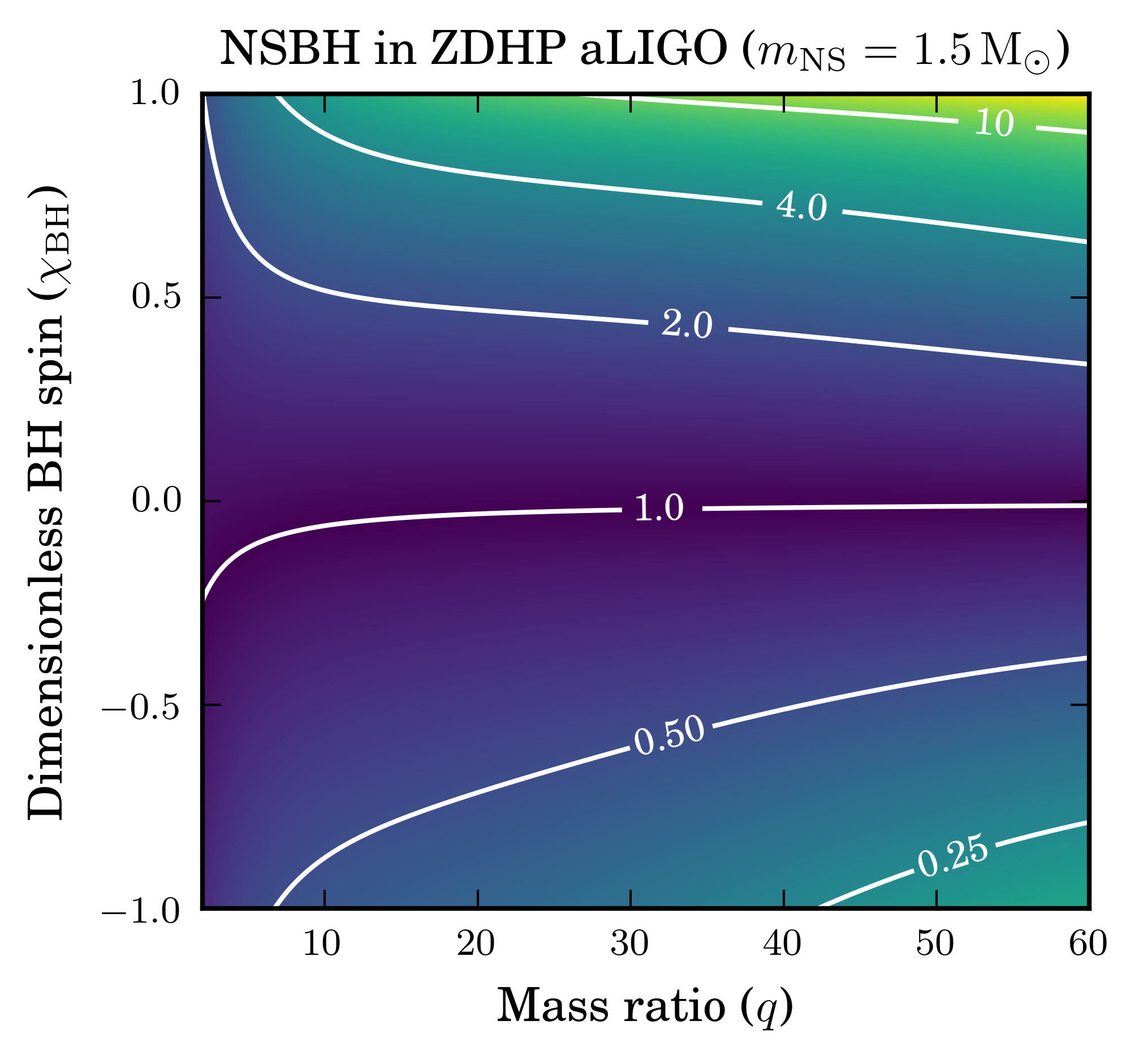} \\
	\includegraphics[width=0.4\textwidth]{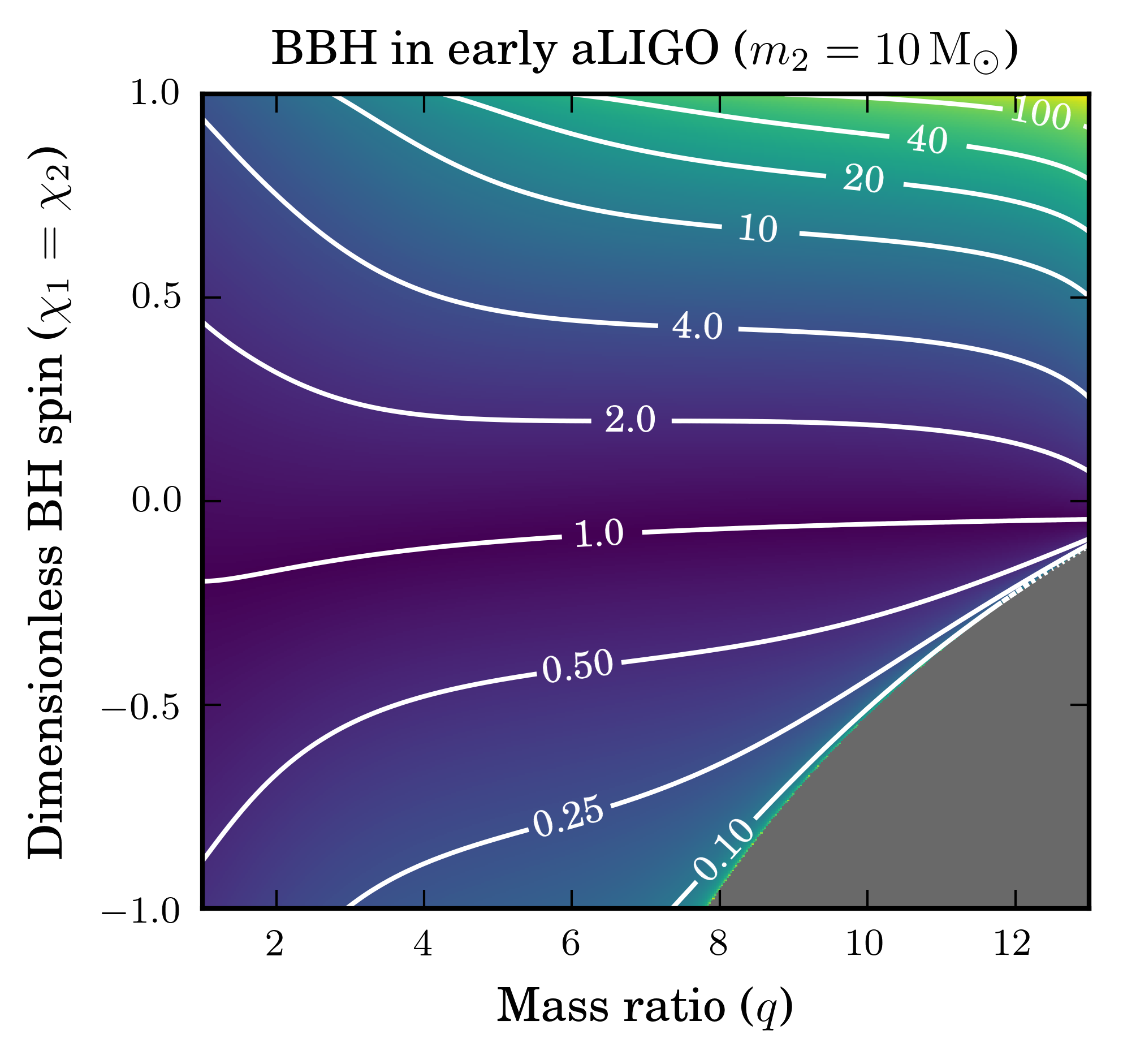} \qquad
	\includegraphics[width=0.4\textwidth]{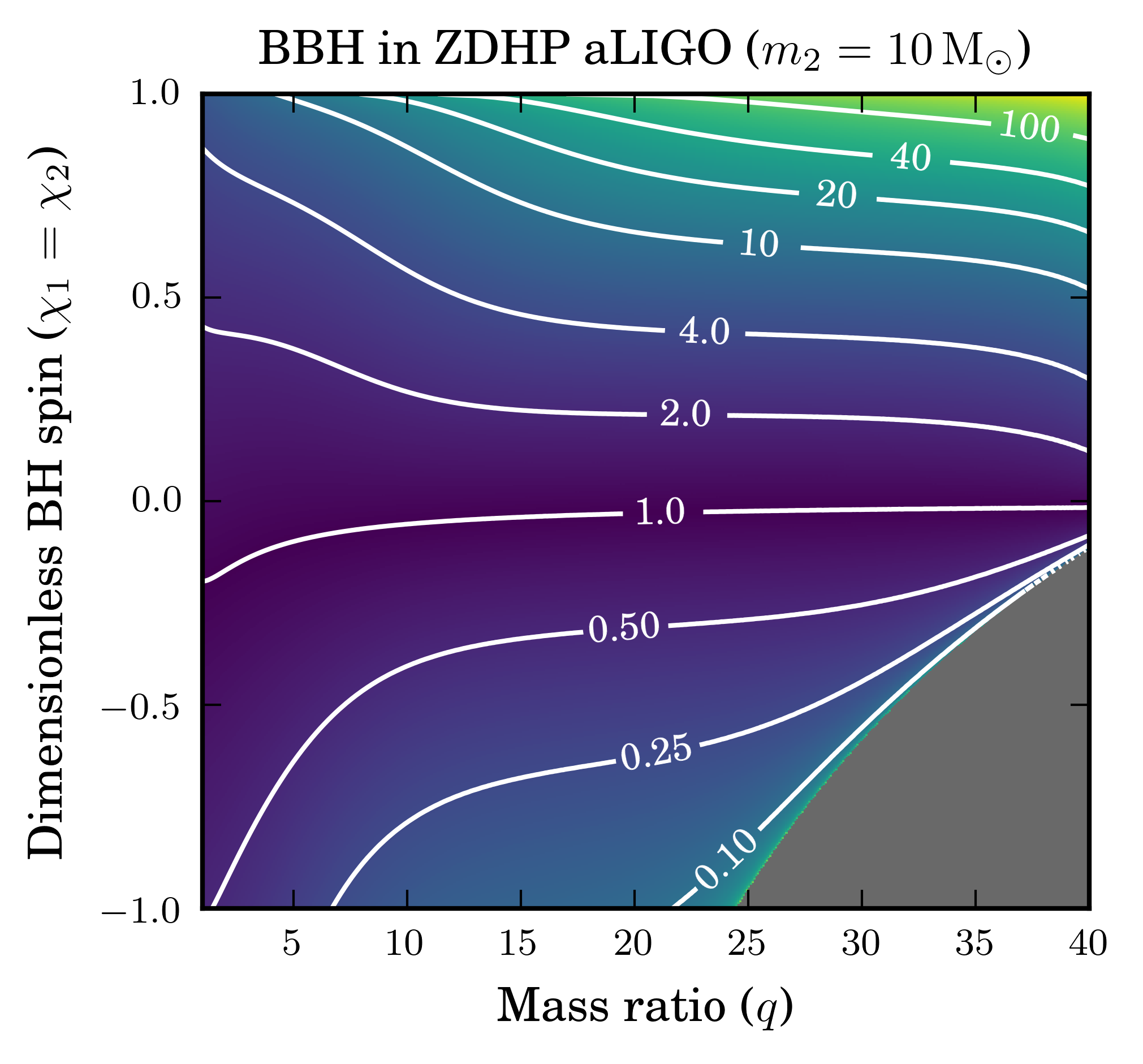}
\caption{Relative SNR between inspiral signals terminating at the hybrid MECO and terminating at the Schwarzschild ISCO as a function of the spin of the black hole(s) and the mass-ratio $q$ of the two objects. The mass of one body is fixed: $m_\textrm{NS} = 1.5 M_{\odot}$ in the NSBH case and $m_2 = 10 M_{\odot}$ in the BBH case. The mass of the other body varies as $m_1 = q m_2$. For early aLIGO the low frequency sensitivity limit is chosen as $f_0 = 30 \mathrm{Hz}$, and for ZDHP aLIGO, $f_0 = 10 \mathrm{Hz}$. Grey regions correspond to systems where the termination frequency lies below $f_0$.}
\label{SNRloss}
\end{figure*}

The optimal SNR of a waveform $h(t)$ is given by~\cite{CutlerFlanagan, PoissonWill}
\begin{equation}
\rho_{\text{opt}} (h) = \sqrt{\langle h | h \rangle} \, .
\end{equation}
Here, $\langle h_1 | h_2 \rangle$ is the noise-weighted inner product defined by
\begin{equation}
\langle h_1 | h_2 \rangle = 2 \int_0^\infty \frac{\tilde{h}_1^*(f) \tilde{h}_2(f) + \tilde{h}_1(f) \tilde{h}_2^*(f)}{S_n(f)} \: \mathrm{d}f \, ,
\end{equation}
where $\tilde{h}_i(f)$ is the Fourier transform of $h_i(t)$ and $^*$ indicates complex conjugation. $S_n(f)$ denotes the one-sided power spectral density of the detector's noise. When restricting attention to only the inspiral part of a signal, the upper frequency cutoff of the integral is the end of the inspiral (or the Nyquist frequency if the latter is reached first). The lower limit of the integral can be taken as the detector's effective low-frequency sensitivity limit, below which the detector is dominated by seismic noise.
The relative SNR between inspiral waveforms terminating at the hybrid MECO, $h_{\text{hM}}$, and waveforms terminating at the Schwarzschild ISCO, $h_{\text{Schw}}$ is therefore
\begin{equation}
\rho_{\text{r}} = \frac{ \rho_{\text{opt}} (h_{\text{hM}}) }{ \rho_{\text{opt}} (h_{\text{Schw}}) }= \sqrt{ \frac{ \langle h_{\text{hM}} | h_{\text{hM}} \rangle }{ \langle h_{\text{Schw}} | h_{\text{Schw}} \rangle} }  \, .
\end{equation}

The Fourier domain waveform in the stationary phase approximation is given by~\cite{PoissonWill}
\begin{equation} 
\tilde{h}(f) = \mathcal{A} f^{-7/6} e^{i \Psi(f)} \, ,
\end{equation}
where $\Psi(f)$ is the phase of the Fourier domain waveform and $\mathcal{A}$ its amplitude.
With this expression, neglecting all the post-Newtonian corrections to the amplitude, the relative SNR between both templates is given by
\begin{equation}
\rho_{\text{r}} = \sqrt{ \frac{ \int_{f_0}^{f_\textrm{hM}} \frac{f^{-7/3}}{S_n(f)} \: \mathrm{d}f }{ \int_{f_0}^{f_\textrm{Schw}} \frac{f^{-7/3}}{S_n(f)} \: \mathrm{d}f } }
\end{equation}
where $f_0$ is the detector's low-frequency sensitivity, $f_\textrm{hM}$ is the frequency of the hybrid MECO, and $f_\textrm{Schw}$ is the frequency of the Schwarzschild ISCO. 

Figure~\ref{SNRloss} shows this relative SNR as a function of the mass ratio $q$. The sensitivity curves used represent
\begin{inparaenum}[(i)]
\item the early runs of the Advanced LIGO generation (which we call early aLIGO), and 
\item the modelled sensitivity of the zero-detuned, high-power design of the mature Advanced LIGO (which we call ZDHP aLIGO)~\cite{aLIGO}. 
\end{inparaenum}
The relative SNR is $\rho_r = 1$ for waveforms with $f_{\text{hM}} \simeq f_\textrm{Schw}$. Systems where $f_{\text{hM}} < f_\textrm{Schw}$ ($f_{\text{hM}} > f_\textrm{Schw}$) have $\rho_r < 1$ ($\rho_r > 1$). 

For the NSBH systems, small mass-ratios give a total mass $M < 10 M_{\odot}$. At a given velocity, the frequency of the gravitational wave is inversely proportional to the total mass of the system, $f_\textrm{GW} = v^3 / (\pi M)$. For low mass systems, therefore, the Schwarzschild ISCO and the hybrid MECO have frequencies $f_\textrm{GW} \gtrsim 400$ Hz. Since the detectors are most sensitive below 400 Hz, changing the waveform cutoff at such high frequencies does not result in a significant gain or loss in the SNR. 

For the BBH systems, however, total masses are higher and the late inspiral frequencies lie in the region where the detectors are most sensitive. This translates into a significant SNR content between the Schwarzschild ISCO and the hybrid MECO frequencies. For sufficiently large mass ratios and positive spin values ($\chi > 0.5$), most of the inspiral signal is contained between the Schwarzschild ISCO and hybrid MECO frequencies. In the case of anti-aligned spins however, the Schwarzschild ISCO actually extends beyond the hybrid MECO and thus contains spurious inspiral signal that is unlikely to be physical.

% -------------------------------------------------------------------------------------------------------------------------------

\section{Application to post-Newtonian approximants}\label{Approximants}

The PN energy, $E^\textrm{PN}$, and energy flux, $\mathcal{F}^\textrm{PN}$, can be used to compute the evolution of the gravitational-wave phase~\cite{BIOPS,T2T4} in quasi-circular inspirals. From the energy balance equation $dE/dt=-\mathcal{F}$, one can obtain an expression for the time evolution of the PN velocity parameter $v$:
\begin{equation} \label{balance}
\frac{dv}{dt} = - \frac{\mathcal{F}(v)}{E'(v)} \, ,
\end{equation}
where $E'(v) = dE/dv$. The gravitational wave frequency $f_\textrm{GW}$ is related to the PN velocity parameter by $v = (\pi M f_\textrm{GW})^{1/3}$. The gravitational-wave phase is twice the orbital phase $\phi(t)$, which is given by
\begin{equation} \label{phase}
\frac{d\phi}{dt} = \frac{v^3}{M} \, .
\end{equation}
The energy balance equation can be solved in different ways to obtain different PN approximants for the gravitational-wave phase. In this paper, we focus on the so-called TaylorT4~\cite{TaylorT4} and TaylorT2 approximants.

The TaylorT4 approximant is obtained by expanding the ratio $\mathcal{F}(v)/E'(v)$ to the consistent PN order and then integrating Eq.~\eqref{balance} numerically. Introducing the resulting $v(t)$ into Eq.~\eqref{phase}, the phase of the gravitational wave can be integrated.

The TaylorT2 approximant instead expands the ratio $E'(v)/\mathcal{F}(v)$ to the consistent PN order: 
\begin{equation} \label{T2}
\frac{dt}{dv} = - \frac{E'(v)}{\mathcal{F}(v)} \, .
\end{equation}
The phase can then be obtained combining~\eqref{phase} and~\eqref{T2}, and integrating
\begin{equation}
\frac{d\phi}{dv} = - \frac{v^3}{M} \frac{E'(v)}{\mathcal{F}(v)} \, .
\end{equation}

Ideally, the phase is integrated up to the MECO frequency in both approximants. However, as discussed above, there are regions of the compact binary parameter space where the traditional MECO does not exist. One wants to avoid integrating beyond the range of validity of the PN approximation. Therefore, further cutoff conditions are imposed~\cite{T2T4}:
\begin{inparaenum}[(i)]
\item the rate of increase in frequency must not decrease ($dv/dt \geq 0$ for TaylorT4), and
\item analogous to the previous condition, $dt/dv \geq 0$ for TaylorT2.
\end{inparaenum}
Consequently, the waveform might terminate before reaching the MECO (when it exists) due to these extra cutoff conditions.

\begin{figure}[tb]
\centering
	\includegraphics[width=\columnwidth]{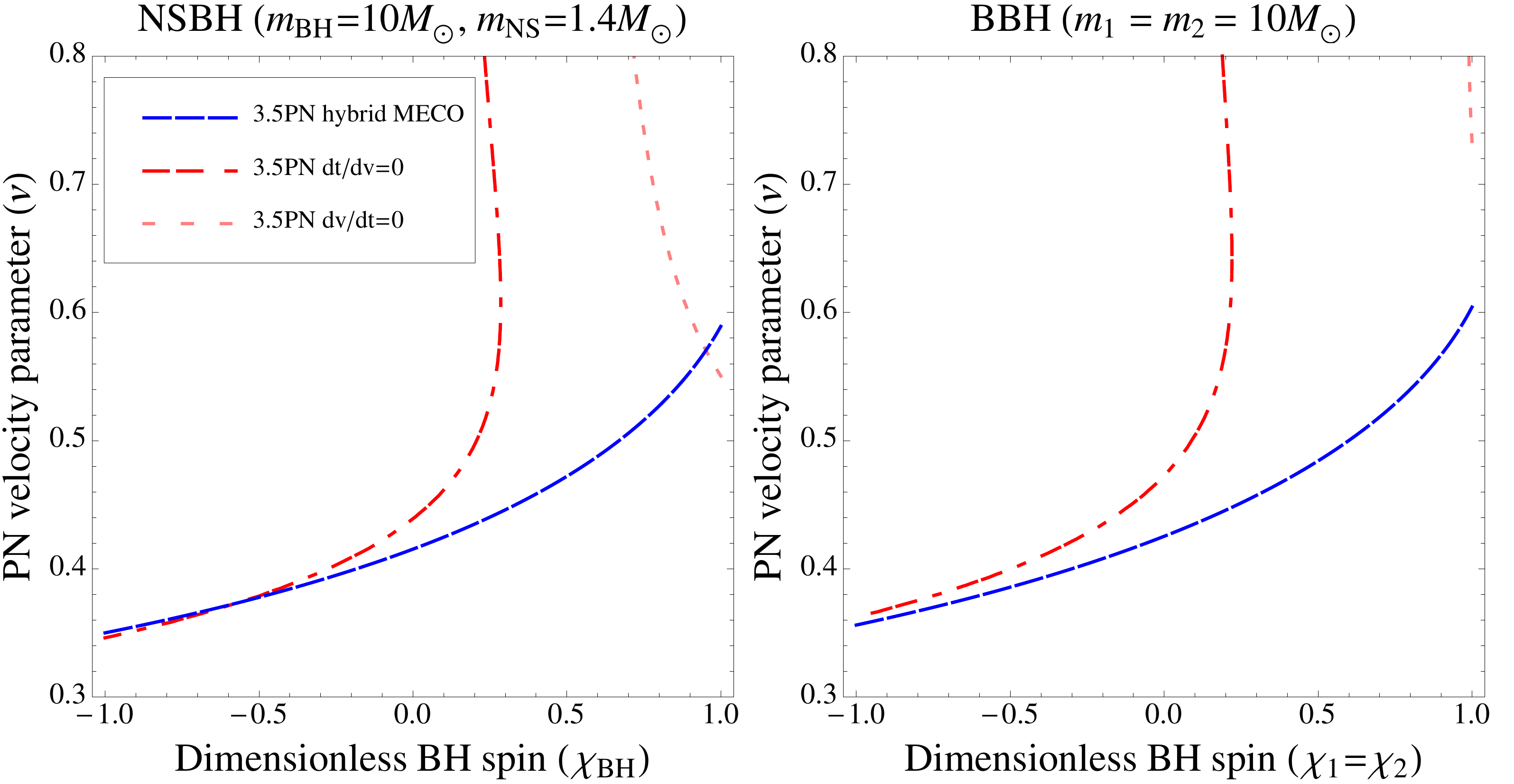}
\caption{Hybrid MECO and the integrands cutoffs at the 3.5PN order. In the NSBH case, TaylorT4 breaks before the hybrid MECO at high aligned spins, and TaylorT2 at high anti-aligned spins. In the BBH case, however, the hybrid MECO is sufficient condition.}
\label{dtdv}
\end{figure}

The hybrid MECO proposed in section~\ref{Corrected_MECO} exists in any region of the parameter space. However, the approximant could still terminate before reaching the hybrid MECO if the cutoff conditions (i) and (ii) are met. Figure~\ref{dtdv} shows the regions of the parameter space where the integrands become zero before reaching the hybrid MECO when including terms up to and including 3.5PN order. In the NSBH case, the $dv/dt=0$ condition is needed at spins above 0.9 for the TaylorT4 approximant, and the $dt/dv=0$ condition is needed at spins below $\chi \simeq -0.5$ for the TaylorT2 approximant. Therefore, the use of the extra cutoff conditions will, in some occasions, still be needed.

% -------------------------------------------------------------------------------------------------------------------------------

\section{Including additional physical effects to the hybrid energy}\label{Tidal}

The simplicity of the hybrid energy defined in Eq.~\eqref{NewEnergy} allows for the easy addition of physical effects not included in the point-mass PN energy or the exact Kerr test-mass limit. For instance, one could consider the effects of internal structure~\cite{referee} or the effect of self-force corrections~\cite{referee2} to see how the hybrid MECO varies under these effects. As an explicit example, here we show the variation in the end of the inspiral due to the possible tidal deformation of neutron stars. 

The dynamics of NSBH binaries are influenced by the tidal deformation of the neutron star when the separation between the two bodies decreases~\cite{TidalDisruption, TidalDisruptionProspects, TidalEffects, TidalEffects2}. The magnitude of the deformation depends on the equation of state (EOS) of the neutron star and the masses and spins of the two objects. For comparable masses, the neutron star may completely disrupt before the plunge into the black hole~\cite{TidalDisruption}. In this case, some of the disrupted mass may become unbound~\cite{TidalDisruption} or form an accretion disk~\cite{EM_GRB}. Such systems are of great astrophysical interest, because they may provide an electromagnetic counterpart to gravitational-wave signals (e.g. short gamma-ray bursts~\cite{EM_GRB}).

An analytical model that accounts for the disruption of the neutron star is more difficult to define. Some works have computed leading-order terms that describe tidal deformation~\cite{TidalEffects,TidalEffects2}.
Tidal corrections to the PN energy are a Newtonian effect that scale proportional to the 5PN order. For NSBH, where the neutron star is assumed to be the only deformable body\footnote{The deformability of black holes and the value of their Love numbers is still ongoing work~\cite{Damour, TidaldefBH, TidaldefBHPoisson} although it is expected to be negligible compared to the deformation of neutron stars.}, these effects are given by~\cite{TidalEffects,TidalEffects2}
\begin{align} \label{Tidalterms}
E_\textrm{tidal} = {} & - \frac{1}{2} \eta v^2 \left[ - \frac{9 m_{1}}{m_{2}} \tilde{\lambda}_{2} v^{10} \right. \non \\
& \left. - \frac{11 m_{1}}{2 m_{2}} \left( 3 + \frac{2 m_{2}}{M} + \frac{3 m_{2}^2}{M^2} \right) \tilde{\lambda}_{2} v^{12} + \mathcal{O}(v^{14}) \right] \, , \non \\
\mathcal{F}_\textrm{tidal} = {} &  \frac{32}{5} \eta^2 v^{10} \left[ \left( \frac{18 M}{m_{2}} - 12 \right) \tilde{\lambda}_{2} v^{10} \right. \non \\
& - \frac{M}{28 m_{2}} \left( 704 + 1803 \frac{m_{2}}{M} - 4501 \frac{m_{2}^2}{M^2} \right. \non \\
& \left. \left. \qquad \qquad + 2170 \frac{m_{2}^3}{M^3} \right) \tilde{\lambda}_{2} v^{12} + \mathcal{O}(v^{14}) \right] \, ,
\end{align}
where $\tilde{\lambda}_{2} = \lambda_2 (m_2 / M)^5$, $\lambda_{2} = \frac{2}{3} k_{2} (R / m_2)^5$ is the dimensionless tidal deformability of the neutron star, $k_{2}$ is the Love number of the neutron star, $R$ is its radius, and $m_2$ is its mass~\cite{LoveN_Tanja, LoveN_Damour}. 

\begin{figure}
\centering
\includegraphics[width=0.4\columnwidth]{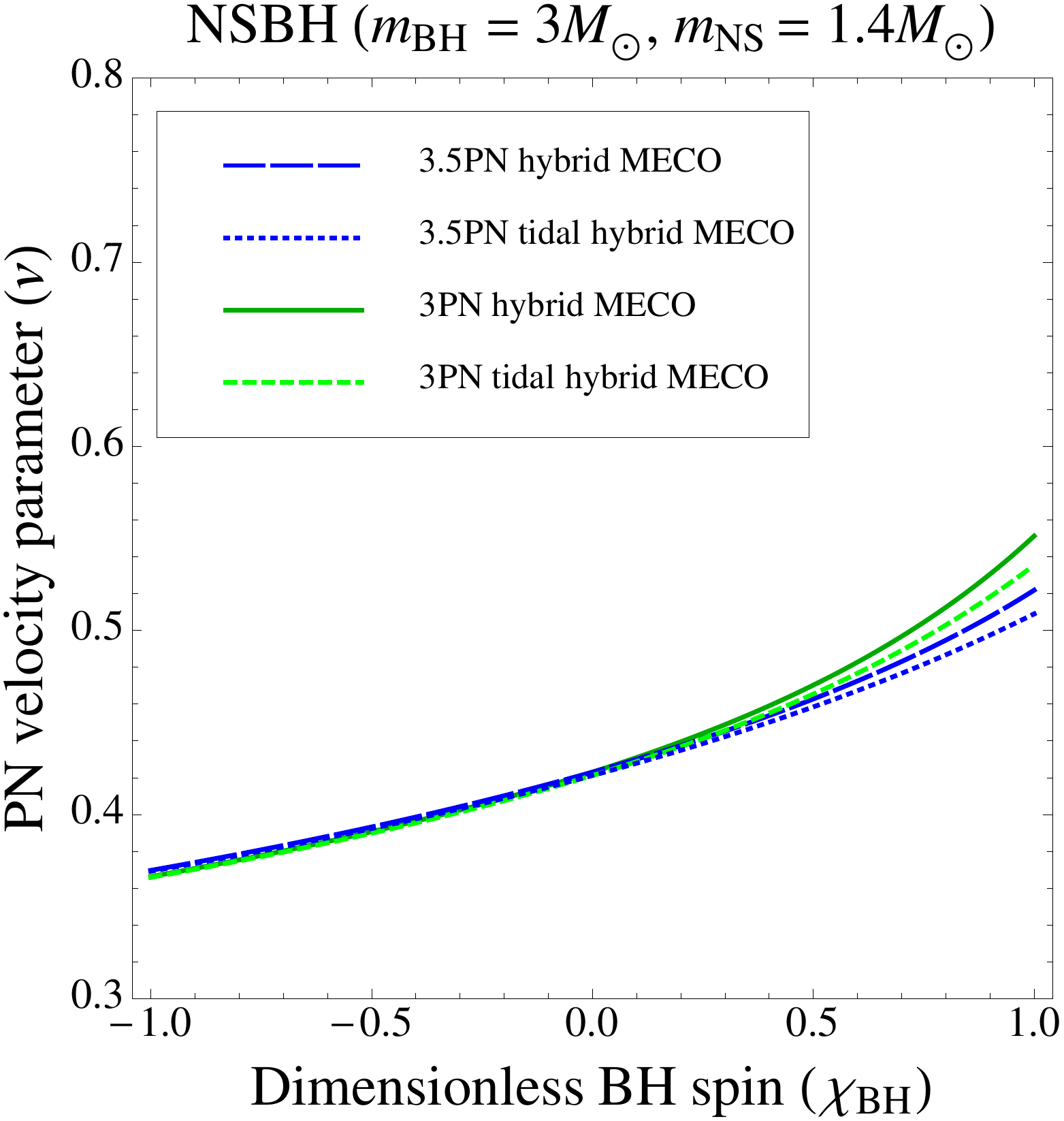}
\includegraphics[width=0.58\columnwidth]{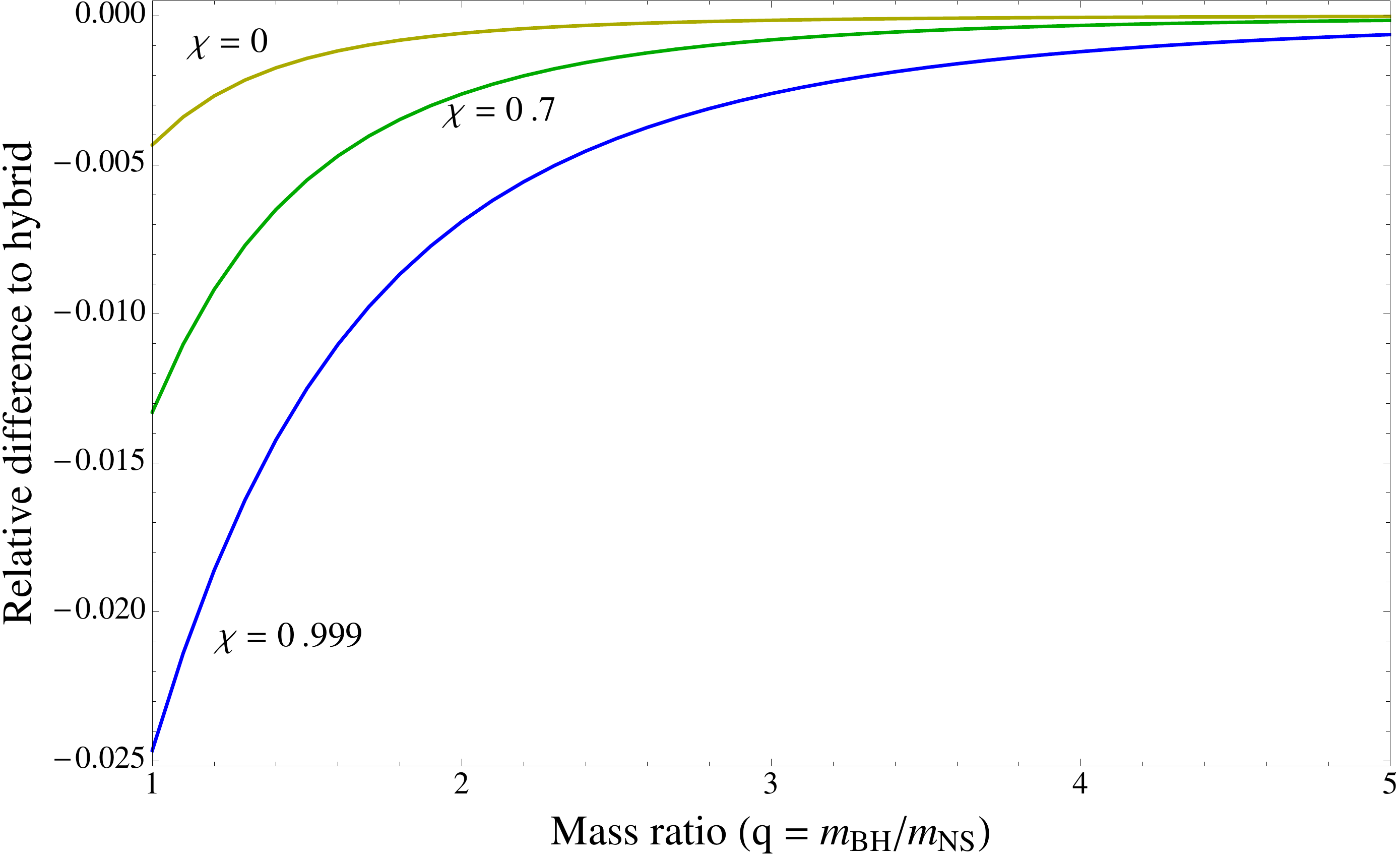}
\caption{Change on the hybrid MECO due to the tidal effects given in Eq.~\eqref{Tidalterms}. The equation of state of the neutron star considered is AP4, which gives a radius $R = 11.5$ km and a dimensionless tidal deformability $\lambda_2 = 269.75$~\cite{LALSuite}. \textit{Left:} qualitative effect of the tidal terms on the hybrid MECO for a mass-ratio $q \simeq 2$. This system is chosen because it clearly shows the effect at high spins. However, total disruption of the neutron star might happen before the MECO is reached. \textit{Right:} relative difference between the hybrid MECO at the 3.5PN order with and without tidal terms $(v_\textrm{tidal} - v_\textrm{hybrid}) / v_\textrm{hybrid}$ for mass-ratios up to $q=5$. The mass of the neutron star is fixed to $m_\textrm{NS} = 1.4 M_{\odot}$ and its spin is set to zero.}
\label{TidalMECO}
\end{figure}

Since energy is absorbed by the tidal effects, the velocity at which the energy reaches its minimum will be smaller when the tidal effects are considered.
Figure~\ref{TidalMECO} shows the change in the hybrid MECO due to the tidal deformation of the neutron star. This deformation hastens the end of the inspiral phase, especially for highly aligned black-hole spins. This effect depends strongly on the EOS and the mass-ratio.

The tidal terms in Eq.~\eqref{Tidalterms} do not account for complete disruption. If the neutron star disrupts before the end of the inspiral, the dynamics of the system change dramatically. If tidal disruption occurs after the end of the inspiral, or not at all, then the dynamics are little changed. Numerical simulations of NSBH~\cite{NR_NSBH, NR_NSBH_spin} provide a better understanding of the disruption mechanism. Based on the resulting numerical waveforms, studies of the end of the inspiral due to the disruption of the neutron star have been performed in~\cite{tidal_disruption}. It is therefore important to determine whether disruption occurs before the tidal hybrid MECO, so Fig.~\ref{TidalMECO} also shows systems that may be tidally disrupted.

% -------------------------------------------------------------------------------------------------------------------------------

\section{Conclusion}

We constructed a hybrid MECO by combining the information of post-Newtonian theory with the exact Kerr solution, and showed that it can be used for binary systems with arbitrary component spins. In combination with NR simulations and full inspiral-merger-ringdown waveform models, this simple analytical approximation gives a more complete understanding of the dynamics of compact binaries.

Unlike the pure PN MECO, which does not exist for large component spins, the hybrid MECO is well-defined for any known PN order in the whole parameter space. This feature is very important for studies of the inspiral phase in real gravitational wave signals, where component spins can have a significant effect. Furthermore, the hybrid MECO is well-defined for equal-mass binaries and, by construction, reduces to the Kerr ISCO in the test-mass limit. This is in contrast to the PN ISCO condition in~\cite{iscoFavata}, which is well-defined only in the test-mass limit and differs from the Kerr ISCO at high spins. In addition, the method we propose can be easily updated when new post-Newtonian terms become available. Finite-mass corrections can also be introduced in the hybrid energy, as we have shown, for instance, with the tidal effects of neutron stars.

In this new era of gravitational-wave astronomy, real signals from compact binaries become available to further test the theory of general relativity and properties of black holes. Tests that rely on measurements purely in the inspiral phase, such as measurements of the initial masses and spins, will require a well motivated cutoff condition to identify the end of the inspiral regime. Within the PN approximation scheme, the hybrid MECO proposed in this paper shows several key advantages over traditional choices such as the Schwarzschild ISCO: the hybrid MECO is spin dependent; it exists for all known PN terms and all values of parameter space; further physical effects are readily included; it reduces to the correct extreme mass limit; and occurs before the expected peak emission of the merger.

Furthermore, direct comparisons between the tidal hybrid MECO and the cutoff obtained from total tidal disruption models may provide more information about the end of the inspiral in systems with neutron stars. This will be important for binaries with tidally deformed or disrupted neutron stars. In this paper we have worked only with component spins aligned or anti-aligned with the orbital angular momentum. However, the behaviour of the MECO could vary significantly for precessing systems. In the future, the hybrid MECO shall be extended to include spin-precessing terms. 

% -------------------------------------------------------------------------------------------------------------------------------

\section*{Acknowledgements}
We are thankful to Ofek Birnholtz, Tito Dal Canton, Thomas Dent, Marc Favata, Evan Goetz, Ian Hinder, Badri Krishnan, Alex Nitz, Frank Ohme, Francesco Pannarale, Jan Steinhoff, and Andrea Taracchini for useful comments and discussions.

% -------------------------------------------------------------------------------------------------------------------------------
% APPENDIX
% -------------------------------------------------------------------------------------------------------------------------------

\appendix

\section{Relations between parameters} \label{Relations}

The Kerr ISCO location given in Eq.~\eqref{Kisco} is coordinate dependent. The orbital radius can be expressed in terms of the PN velocity parameter $v$, which is related to the frequency of gravitational waves by $v = (\pi M f_\textrm{GW})^{1/3}$, thus obtaining a gauge-invariant ISCO independent of the total mass of the system:
\begin{equation}
\frac{M}{r} = \frac{v^2}{(1 - \chi v^3)^{2/3}} \, .
\end{equation}
At zero spin, this relation reduces to $v = \sqrt{M/r}$, which gives a velocity of $v = 1 / \sqrt{6} \simeq 0.41$ for the Schwarzschild ISCO. For extreme spins ($a = m$), the Kerr ISCO corresponds to velocities of $v \simeq 0.79$ (aligned case) and $v \simeq 0.34$ (anti-aligned case).

One can use the phase of the gravitational wave to compute the number of cycles between different terminations.
The number of gravitational-wave cycles $\mathcal{N}$ between the velocities $v_1$ and $v_2$ can be given by
\begin{equation}
\mathcal{N} = \frac{\phi^\textrm{GW} (v_2) - \phi^\textrm{GW} (v_1)}{2 \pi} \, ,
\end{equation}
where $\phi^\textrm{GW} (v)$ is the gravitational-wave phase (twice the orbital phase). The gravitational-wave phase as function of the frequency can be obtained from equation (3.11) in~\cite{CutlerFlanagan} (or the orbital phase as function of the velocity from equations (318) and (3.8a) in~\cite{BlanchetPN2014} and~\cite{BIOPS}, respectively). At the leading order, the gravitational-wave phase is independent of the spin and is given by
\begin{equation}
\phi^\textrm{GW} (v) = - \frac{1}{16 v^5 \eta} \, .
\end{equation}
Between the minimum (anti-aligned) and the maximum (aligned) velocity values of the Kerr ISCO, the number of gravitational-wave cycles for an equal-mass binary ($\eta = 0.25$) is $\mathcal{N} \simeq 8.6$, and $\mathcal{N} \simeq 43.1$ for a binary with $\eta = 0.05$. In the bandwidth of a gravitational-wave detector with lower frequency cutoff at $f_0 = 30$ Hz, the gravitational wave of an equal-mass BBH with total mass $M = 20 M_{\odot}$ will be visible from $v_1 \simeq 0.21$ and will show $\mathcal{N} \simeq 97.3$ cycles when it reaches $v_2 \simeq 0.79$. For this gravitational wave, the 8.6 cycles between the Kerr ISCO velocities in the equal-mass case represent the $8.8\%$ of the inspiral phase. 

\section{Post-Newtonian energy and energy flux} \label{PNformulae}

\newcommand{\kp}{\kappa_{+}}
\newcommand{\km}{\kappa_{-}}
\newcommand{\lp}{\lambda_{+}}
\newcommand{\lm}{\lambda_{-}}
\newcommand{\SL}{S_\textrm{L}}
\newcommand{\SigL}{\Sigma_\textrm{L}}

The derivation of the different terms of the post-Newtonian energy and energy flux in the centre-of-mass frame can be found in the literature~\cite{BlanchetPN2014, 4PN, spin-orbit, LeviSteinhoff_spin-orbit, 2PNspin-spin, 3PNspin-spin, LeviSteinhoff_3PNspin-spin, LeviSteinhoff_4PNspin-spin, spin-cubed}. This Appendix is a collection, in one place, of all the terms used in this paper. Since the orbits of compact binaries are expected to circularise by the time they become visible with current gravitational-wave detectors, we restrict ourselves to the expressions for circular orbits. The different terms in the energy per unit total mass and the energy flux are:
\begin{align}
E^\textrm{PN} = {} & - \frac{1}{2} \eta v^2 \left[ E_\textrm{NS} + E_\textrm{SO} + E_\textrm{SS} + E_{\textrm{S}^3} + E_{\textrm{S}^4} \right] \, , \label{PNenergy} \\
\mathcal{F}^\textrm{PN} = {} & \frac{32}{5} \eta^2 v^{10} \left[ \mathcal{F}_\textrm{NS} + \mathcal{F}_\textrm{SO} + \mathcal{F}_\textrm{SS} + \mathcal{F}_{\textrm{S}^3} \right] \, , \label{PNflux}
\end{align}
where NS are the non-spinning terms, SO the spin-orbit terms, SS the spin-spin terms, $S^3$ the spin-cubed terms, and $S^4$ the spin-quartic terms. The individual masses of the bodies are denoted $m_1$ and $m_2$, $M = m_1 + m_2$ is the total mass of the binary, the $v$ parameter is given by the sum of the speeds in the rest-mass frame $v=v_{1}+v_{2}$, and $\eta = m_{1} m_{2} / M^2$ is the symmetric mass ratio.

The non-spinning energy terms are known up to 4PN~\cite{BIOPS,BlanchetPN2014}
\begin{widetext}
\begin{align} \label{nonspinPNenergy}
E_\textrm{NS} = {} & 1 - \left( \frac{3}{4} + \frac{\eta}{12} \right) v^2 - \left( \frac{27}{8} - \frac{19}{8} \eta + \frac{\eta^2}{24} \right) v^4 - \left[ \frac{675}{64} - \left( \frac{34445}{576} - \frac{205 \pi^2}{96} \right) \eta + \frac{155}{96} \eta^2 + \frac{35}{5184} \eta^3 \right] v^6 \non \\
& - \bigg[ \frac{3969}{128} + \left( \frac{123671}{5760} - \frac{9037 \pi^2}{1536} - \frac{1792}{15}  \ln 2 - \frac{896}{15} \gamma_\textrm{E} \right) \eta + \left( \frac{498449}{3456} - \frac{3157 \pi^2}{576} \right) \eta^2 \non \\
& \qquad - \frac{301}{1728} \eta^3 - \frac{77}{31104} \eta^4 - \frac{448}{15} \eta \ln v^2 \bigg] v^8 + \mathcal{O} \left( v^{10} \right) \, ,
\end{align}
\end{widetext}
where $\gamma_\textrm{E}$ is the Euler constant.

\clearpage
For the spin-orbit terms, we define the variables
\begin{equation} \label{SOPNenergy}
\begin{aligned}
\mathbf{S} & = \mathbf{S}_{1} + \mathbf{S}_{2} \, , 	&	 \qquad \mathbf{\Sigma} & = M \left( \frac{\mathbf{S}_{2}}{m_{2}} - \frac{\mathbf{S}_{1}}{m_{1}} \right) \, , \non \\
S_\textrm{L }& = \frac{\mathbf{S} \cdot \mathbf{L}}{M^2} \, , 	&	 \qquad \SigL & = \frac{\mathbf{\Sigma} \cdot \mathbf{L}}{M^2}  \, , 
\end{aligned}
\end{equation}
where $\mathbf{S}_{1}$ and $\mathbf{S}_{2}$ are the individual vector spins of the objects, and $\mathbf{L}$ is the unit vector pointing in the direction of the orbital angular momentum. We will also use $S_\textrm{1L} = \mathbf{S}_{1} \cdot \mathbf{L}$ and $S_\textrm{2L} = \mathbf{S}_{2} \cdot \mathbf{L}$. 

The spin-orbit terms are known up to 3.5PN \cite{spin-orbit, LeviSteinhoff_spin-orbit}
\begin{align}
E_\textrm{SO} = & \left( \frac{14}{3} \SL + 2 \delta \SigL \right) v^3 \non \\
& + \left[ \left( 11 - \frac{61}{9} \eta \right) \SL + \left( 3 - \frac{10}{3} \eta \right) \delta \SigL \right] v^5 \non \\
& + \left[ \left( \frac{135}{4} - \frac{367}{4} \eta + \frac{29}{12} \eta^2 \right) \SL \right. \non \\
& \left. \qquad + \left( \frac{27}{4} - 39 \eta + \frac{5}{4} \eta^2 \right) \delta \SigL \right] v^7 + \mathcal{O} \left( v^{8} \right) \, ,
\end{align}
where $\delta = (m_{1} - m_{2}) / M$.

For the following spinning terms, we restrict ourselves to the expressions for spin-aligned binaries, and introduce the quantity $q_i = m_i / M$.
 
The 2PN spin-spin term is given by \cite{2PNspin-spin, LeviSteinhoff_3PNspin-spin}
\begin{equation}
E_\textrm{SS}^\textrm{2PN} = - \frac{1}{M^2} \left( \frac{2 S_1 S_2}{m_1 m_2} + \frac{S_1^2 \kappa_1}{m_1^2} + \frac{S_2^2 \kappa_2}{m_2^2} \right) \, ,
\end{equation}
where the constants $\kappa_i$ are the quadrupole terms that represent the distortion of the bodies due to their spins~\cite{qm}. For black holes, $\kappa_i = 1$. For neutron stars, $\kappa_i \simeq 4 - 8$ depending on the equation of state.

The 3PN spin-spin term is given by~\cite{3PNspin-spin, LeviSteinhoff_3PNspin-spin}
\begin{align}
E_\textrm{SS}^\textrm{3PN} = - \frac{1}{M^2} & \Bigg\{ \frac{5}{9} \frac{S_1 S_2}{m_1 m_2} ( 3 + \eta ) \non \\
& - \frac{S_1^2}{m_1^2} \left[ \frac{5}{9} \left( 9 - 3 \eta - 5 q_1^2 \right) \right. \non \\
& \qquad \qquad \left. - \frac{5 \kappa_1}{6} \left( 3 + 3 \eta + 4 q_1^2 \right) \right] \non \\
& - \frac{S_2^2}{m_2^2} \left[ \frac{5}{9} \left( 9 - 3 \eta - 5 q_2^2 \right) \right. \non \\
& \qquad \qquad \left. - \frac{5 \kappa_2}{6} \left( 3 + 3 \eta + 4 q_2^2 \right) \right] \Bigg\} \, .
\end{align}

The 4PN spin-spin term is given by~\cite{LeviSteinhoff_4PNspin-spin}
\begin{align}
E_\textrm{SS}^\textrm{4PN} = {} & - \frac{7}{18 M^2} \left[ \frac{S_1 S_2}{12 m_1 m_2} ( 135 - 429 \eta - 53 \eta^2 ) \right. \non \\
& + \frac{S_1^2}{7 M^2} \left( \frac{1}{3} (360 - 749 \eta) + \kappa_1 (279 - 79 \eta) \right) \non \\
& - \frac{S_1^2}{m_1^2} \left( 27 + 6 \eta + 31 \eta^2 - \frac{3 \kappa_1}{4} (27 + 11 \eta - 13 \eta^2) \right) \non \\
& + \frac{S_2^2}{7 M^2} \left( \frac{1}{3} (360 - 749 \eta) + \kappa_2 (279 - 79 \eta) \right) \non \\
& \left. - \frac{S_2^2}{m_2^2} \left( 27 + 6 \eta + 31 \eta^2 - \frac{3 \kappa_2}{4} (27 + 11 \eta - 13 \eta^2) \right) \right] \, .
\end{align}

The spin-cubed term is given by~\cite{spin-cubed, LeviSteinhoff_spin-orbit}
\begin{align}
E_{\textrm{S}^3} = - \frac{2}{M^3} & \Big\{ \frac{S_1^3}{m_1^3} \left[ (3 - q_1) \kappa_1 - 2 \lambda_1 \right] \non \\
& + \frac{S_1^2 S_2}{m_1^2 m_2} \left[ 6 - 2 q_1 - (4 - q_1) \kappa_1 \right] \non \\
& + \frac{S_1 S_2^2}{m_1 m_2^2} \left[ 6 - 2 q_2 - (4 - q_2) \kappa_2 \right] \non \\
& + \frac{S_2^3}{m_2^3} \left[ (3 - q_2) \kappa_2 - 2 \lambda_2 \right] \, ,
\end{align}
where the constants $\lambda_i$ are the octupole terms that characterise the deformation of the bodies due to their spins~\cite{spin-cubed}. For black holes, $\lambda_i = \kappa_i = 1$. For neutron stars, the value of $\lambda_i$ is yet unknown.

The spin-quartic term is given by~\cite{LeviSteinhoff_4PNspin-spin}
\begin{align} \label{S4}
E_{\textrm{S}^4} =  - \frac{7}{M^4} & \Big\{ \frac{S_1^2 S_2^2}{m_1^2 m_2^2} (\kappa_1 \kappa_2 - 1)  \non \\
& + \frac{1}{4} \frac{S_1^4}{m_1^4} (C_1 - \kappa_1^2) + \frac{S_1^3 S_2}{m_1^3 m_2} (\lambda_1 - \kappa_1) \non \\
& + \frac{1}{4} \frac{S_2^4}{m_2^4} (C_2 - \kappa_2^2) + \frac{S_1 S_2^3}{m_1 m_2^3} (\lambda_2 - \kappa_2) \Big\} \, ,
\end{align}
where the constants $C_i$ are the hexadecapole terms that characterise the deformation of the bodies due to their spins~\cite{LeviSteinhoff_4PNspin-spin}. Note that this spin-quartic term vanishes for binary black hole systems ($\kappa_i = \lambda_i = C_i = 1$) and for the neutron-star black-hole binaries we consider in this paper ($S_2 = S_{NS} = 0$).

The non-spinning energy flux terms are known up to 3.5PN~\cite{BIOPS,BlanchetPN2014}
\begin{widetext}
\begin{align}
\mathcal{F}_\textrm{NS} =  {} & 1 - \left( \frac{1247}{336} + \frac{35}{12} \eta \right) v^2 + 4 \pi v^3 - \left( \frac{44711}{9072} - \frac{9721}{504} \eta - \frac{65}{18} \eta^2 \right) v^4 - \left(\frac{8191}{672} + \frac{538}{24} \eta \right) \pi v^5 \non \\
& + \bigg[ \frac{6643739519}{69854400} + \frac{16 \pi^2}{3} - \frac{1712}{105} \gamma_\textrm{E} - \left( \frac{134543}{7776} - \frac{41 \pi^2}{48} \right) \eta - \frac{94403}{3024} \eta^2 - \frac{775}{324} \eta^3 - \frac{856}{105} \ln\left[16v^2\right] \bigg] v^6 \non \\
& - \left( \frac{16285}{504} - \frac{214745}{1728} \eta - \frac{193385}{3024} \eta^2 \right) \pi v^7 + \mathcal{O} \left( v^{8} \right) \, ,
\end{align}
\end{widetext}
where $\gamma_\textrm{E}$ is the Euler constant.

The spin-orbit terms are known up to 3.5PN~\cite{spin-orbit}
\begin{align}
\mathcal{F}_\textrm{SO} = {} & - \left( 4 \SL + \frac{5}{4} \delta \SigL \right) v^3 \non \\
& + \left[ \left( - \frac{9}{2} + \frac{272}{9} \eta \right) \SL + \left( - \frac{13}{16} + \frac{43}{4} \eta \right) \delta \SigL \right] v^5 \non \\
& - \left( 16 \pi \SL + \frac{31 \pi}{6} \delta \SigL \right) v^6 \non \\
& + \left[ \left( \frac{476645}{6804} + \frac{6172}{189} \eta - \frac{2810}{27} \eta^2 \right) \SL \right. \non \\
& \left. \quad + \left( \frac{9535}{336} + \frac{1849}{126} \eta - \frac{1501}{36} \eta^2 \right) \delta \SigL \right] v^7 + \mathcal{O} \left( v^{8} \right) ,
\end{align}
where $\delta = (m_{1} - m_{2}) / M$.

The 2PN spin-spin term is given by~\cite{2PNspin-spin, 3PNspin-spin} 
\begin{align}
\mathcal{F}_\textrm{SS}^\textrm{2PN} = {} & \left[ \SL^2 \left( 2 \kp + 4 \right) + \SL \SigL \left( 2 \delta \kp + 4 \delta - 2 \km \right) \right. \non \\
& \left. + \SigL^2 \left( \left( - \delta \km + \kp + \frac{1}{16} \right) + \eta \left( -2 \kp -4 \right) \right) \right] v^4 \, ,
\end{align}
where $\kp = \kappa_{1} + \kappa_{2}$ and $\km = \kappa_{1} - \kappa_{2}$.

The 3PN spin-spin term is given by~\cite{3PNspin-spin}
\begin{align}
\mathcal{F}_\textrm{SS}^\textrm{3PN} = {} & \bigg\{ \SL^2 \left[ \left( \frac{41 \delta \km}{16} - \frac{271 \kp}{112} - \frac{5239}{504} \right) - \eta \left( \frac{43 \kp}{4} + \frac{43}{2} \right) \right] \non \\
& + \SL \SigL \left[ - \left( \frac{279 \delta \kp}{56} + \frac{817 \delta}{56} - \frac{279 \km}{56} \right) \right. \non \\
& \left. \qquad \qquad - \eta \left( \frac{43 \delta \kp}{4} + \frac{43 \delta}{2} + \frac{\km}{2} \right) \right] \non \\
& + \SigL^2 \left[ \left( \frac{279 \delta \km}{112} - \frac{279 \kp}{112} - \frac{25}{8} \right) \right. \non \\
& \qquad \qquad + \eta \left( \frac{45 \delta \km}{16} + \frac{243 \kp}{112} + \frac{344}{21} \right) \non \\
& \left. \qquad \qquad + \eta^2 \left( \frac{43 \kp}{4} + \frac{43}{2} \right) \right] \bigg\} \, v^6 \,. 
\end{align}

The spin-cubed term is given by~\cite{spin-cubed}
\begin{align}
\mathcal{F}_{\textrm{S}^3} = {} & \Big\{ - \SL^3 \left( \frac{16 \kp}{3} +  4 \lp - \frac{40}{3} \right) \non \\
& - \SL^2 \SigL \left( \frac{35 \delta \kp}{6} + 6 \delta \lp - \frac{73 \delta}{3} + \frac{3 \km}{4} - 6 \lm \right) \non \\
& - \SL \SigL^2 \left[ \frac{35 \delta \km}{12} - 6 \delta \lm - \frac{35 \kp}{12} + 6 \lp - \frac{32}{3} \right. \non \\
& \left. \qquad \qquad - \eta \left( \frac{22 \kp}{3} + 12 \lp - \frac{172}{3} \right) \right] \non \\
& + \SigL^3 \left[ \frac{67 \delta \kp}{24} - 2 \delta \lp - \frac{\delta}{8} - \frac{67 \km}{24} + 2 \lm \right. \non \\
& \left. \qquad + \eta \left( \frac{\delta \kp}{2} + 2 \delta \lp - 11 \delta + \frac{61 \km}{12} - 6 \lm \right) \right] \Big\} \, v^7 \, ,
\end{align}
where $\lp = \lambda_{1} + \lambda_{2}$ and $\lm = \lambda_{1} - \lambda_{2}$.
\vfill

\bibliography{Hybrid_MECO}

\end{document}